\documentclass[
reprint,amssymb,
amsmath,nobibnotes,siunitx,aps,prm,
floatfixaps,citeautoscript,showkeys,
floatfix,longbibliography,balance,
superscriptaddress,
bibnotes,prm,
]{revtex4-2}
\usepackage[table,dvipsnames]{xcolor}
\usepackage[utf8]{inputenc}
\usepackage[T1]{fontenc}
\usepackage{graphicx}
\usepackage{grffile}
\usepackage{longtable}
\usepackage{threeparttable}
\usepackage{multirow}
\usepackage{booktabs}
\usepackage{wrapfig}
\usepackage{rotating}
\usepackage[normalem]{ulem}
\usepackage{amsmath}
\usepackage{textcomp}
\usepackage{amssymb}
\usepackage{times}
\usepackage{capt-of}
\usepackage{hyperref}
\usepackage{lipsum}
\usepackage[]{hyperref}
\hypersetup{colorlinks,pdfborder={0,0,0},linkbordercolor={white},citecolor=blue,filecolor=blue,linkcolor=blue,urlcolor=blue}
\usepackage{pbox}
\usepackage{diagbox}
\usepackage{makecell}
\usepackage[UKenglish,USenglish,english,american]{babel}
\usepackage[]{upgreek}
\usepackage[version=4]{mhchem}

\date{\today}
\begin{document}

\bibliographystyle{apsrev4-2}
\pagenumbering{gobble}

\newcommand{\adisl}{$\langle \bold{a} \rangle $ }
\newcommand{\adislns}{$\langle \bold{a} \rangle \mkern-5mu$ }
\newcommand{\cdisl}{$\langle \bold{c} \rangle $ }
\newcommand{\cadisl}{$\langle \bold{c}+\bold{a} \rangle $ }
\newcommand{\ccadisl}{$\langle 2\bold{c}+\bold{a} \rangle $ }
\newcommand{\ccadislns}{$\langle 2\bold{c}+\bold{a} \rangle \mkern-5mu$ }
\newcommand{\eadisl}{\langle \bold{a} \rangle }
\newcommand{\ecdisl}{\langle \bold{c} \rangle }
\newcommand{\epdisl}{\langle \bold{p} \rangle }
\newcommand{\eqdisl}{\langle \bold{q} \rangle }
\newcommand{\ecadisl}{\langle \bold{c}+\bold{a} \rangle }

\newcommand{\bccadisl}{$\langle 111 \rangle /2 $ }

\newcommand{\angstrom}{\mbox{\normalfont\AA}}
\newcommand{\parall}{\mathbin{\!/\mkern-5mu/\!}}

\newcommand{\Vect}[1]{\mathbf{#1}}
\newcommand{\Tens}[1]{\mathbf{#1}}

\newcommand{\tdiff}[1]{\text{d}{#1}}
\newcommand{\redhl}[1]{\textcolor{red}{#1}}
\newcommand{\greenhl}[1]{\textcolor{Green}{#1}}
\newcommand{\orghl}[1]{\textcolor{orange}{#1}}
\newcommand{\bluehl}[1]{\textcolor{blue}{#1}}

\title{\huge  Machine Learning Moment Tensor Potential for Modelling Dislocation and Fracture in L1\(_\text{0}\)-TiAl and D0\(_\text{19}\)-\ce{Ti3Al} Alloys}

\author{Ji Qi}
\affiliation{Materials Science and Engineering Program, University of California San Diego, 9500 Gilman Dr, Mail Code 0448, La Jolla, CA 92093-0448, United States}
\author{Z.~H. Aitken}
\affiliation{Institute of High Performance Computing (IHPC), Agency for Science, Technology and Research (A*STAR), 1 Fusionopolis Way, Connexis \#16-16, Singapore 138632, Republic of Singapore}
\author{Qingxiang Pei}
\affiliation{Institute of High Performance Computing (IHPC), Agency for Science, Technology and Research (A*STAR), 1 Fusionopolis Way, Connexis \#16-16, Singapore 138632, Republic of Singapore}
\author{Anne Marie Z. Tan}
\affiliation{Institute of High Performance Computing (IHPC), Agency for Science, Technology and Research (A*STAR), 1 Fusionopolis Way, Connexis \#16-16, Singapore 138632, Republic of Singapore}
\author{Yunxing Zuo}
\affiliation{Department of NanoEngineering, University of California San Diego, 9500 Gilman Dr, Mail Code 0448, La Jolla, CA 92093-0448, United States}
\author{M.~H. Jhon}
\affiliation{Institute of High Performance Computing (IHPC), Agency for Science, Technology and Research (A*STAR), 1 Fusionopolis Way, Connexis \#16-16, Singapore 138632, Republic of Singapore}
\author{S.~S. Quek}
\affiliation{Institute of High Performance Computing (IHPC), Agency for Science, Technology and Research (A*STAR), 1 Fusionopolis Way, Connexis \#16-16, Singapore 138632, Republic of Singapore}
\author{T. Wen}
\affiliation{Department of Mechanical Engineering, The University of Hong Kong, Hong Kong SAR, China}

\author{Zhaoxuan Wu}
\email[Corresponding author: ]{zhaoxuwu@cityu.edu.hk}
\affiliation{Department of Materials Science and Engineering, City University of Hong Kong, Hong Kong SAR, China}
\affiliation{Hong Kong Institute for Advanced Study, City University of Hong Kong, Hong Kong SAR, China}

\author{Shyue Ping Ong}
\email[Corresponding author: ]{ongsp@eng.ucsd.edu}
\affiliation{Materials Science and Engineering Program, University of California San Diego, 9500 Gilman Dr, Mail Code 0448, La Jolla, CA 92093-0448, United States}
\affiliation{Department of NanoEngineering, University of California San Diego, 9500 Gilman Dr, Mail Code 0448, La Jolla, CA 92093-0448, United States}

\date{\today}

\begin{abstract}

Dual-phase \(\gamma\)-TiAl and \(\alpha_2\)-\ce{Ti3Al} alloys exhibit high strength and creep resistance at high temperatures.  However, they suffer from low tensile ductility and fracture toughness at room temperature.  Experimental studies show unusual plastic behaviour associated with ordinary and superdislocations, making it necessary to gain a detailed understanding on their core properties in individual phases and at the two-phase interfaces.  Unfortunately, extended superdislocation cores are widely dissociated beyond the length scales practical for routine first-principles density-functional theory (DFT) calculations, while extant interatomic potentials are not quantitatively accurate to reveal mechanistic origins of the unusual core-related behaviour in either phases.  Here, we develop a highly-accurate moment tensor potential (MTP) for the binary Ti-Al alloy system using a DFT dataset covering a broad range of intermetallic and solid solution structures.  The optimized MTP is rigorously benchmarked against both previous and new DFT calculations, and unlike existing potentials, is shown to possess outstanding accuracy in nearly all tested mechanical properties, including lattice parameters, elastic constants, surface energies, and generalized stacking fault energies (GSFE) in both phases.  The utility of the MTP is further demonstrated by producing dislocation core structures largely consistent with expectations from DFT-GSFE and experimental observations.  The new MTP opens the path to realistic modelling and simulations of bulk lattice and defect properties relevant to the plastic deformation and fracture processes in \(\gamma\)-TiAl and \(\alpha_2\)-\ce{Ti3Al} dual-phase alloys.

\end{abstract}

\maketitle
\pagenumbering{arabic}

\section{Introduction}

\begin{figure}[!htbp]
 \centering
 \includegraphics[width=0.5\textwidth]{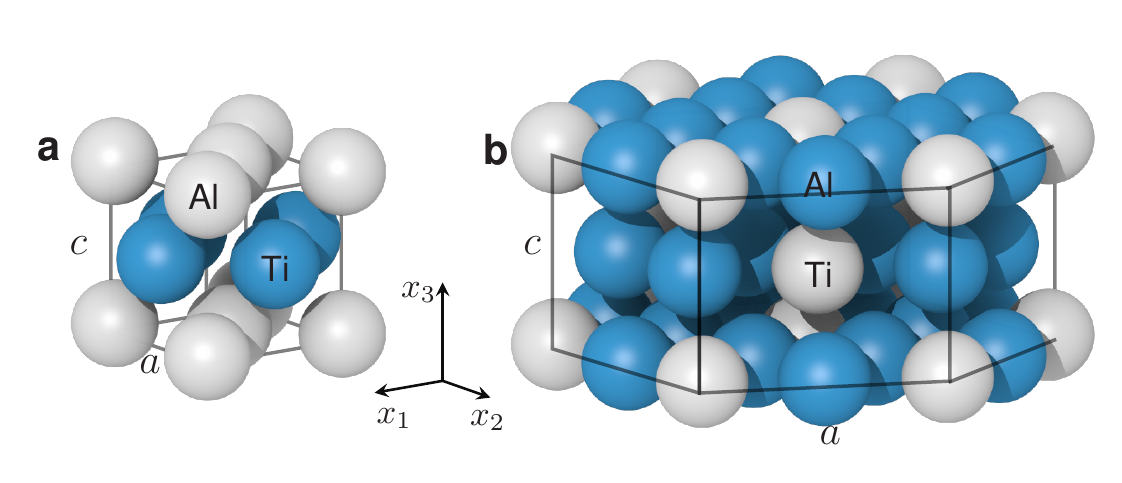}
 \caption{\label{fig:unit_cell_gamma_alpha2} Unit cells of the (a) L1\(_0\) \(\upgamma\)-TiAl and (b) D0\(_{19}\) \(\upalpha_2\)-\ce{Ti3Al} crystal structures. The L1\(_0\) structure is non-cubic with \(c/a \neq 1\). }
\end{figure}

The Ti-Al alloy system possesses superior properties such as light weight and high strength relative to other alloy systems.  In particular, two-phase intermetallic L1\(_\text{0}\) \(\gamma\)-TiAl and D0\(_{19}\) \(\alpha_2\)-\ce{Ti3Al} alloys have outstanding mechanical and physical properties attractive for high-temperature structure material applications~\cite{appel_2011_gtaa}.  They have high specific strength, stiffness and oxidation resistance competitive to other Fe-based or Ni-based alloys~\cite{dimiduk_1999_msea}.  However, TiAl-based alloys suffer from low tensile ductility and fracture toughness at low or room temperatures.  The poor plastic properties stem from the lack of sustainable deformation mechanisms within the individual phases and at the two-phase interfaces~\cite{katzarov_2009_am,palomares_2018_md,ackerman_2020_prm}.  The L1\(_0\) and D0\(_{19}\) structures have highly complex dislocation slip and twinning systems~\cite{chen_2002_msea,chen_2016_natmat} (Fig.~\ref{fig:unit_cell_gamma_alpha2}), and exhibit strong solute interaction effects~\cite{liu_1996_im,yoo_1998_mmta,appel_2016_pms,duan_2022_jac}. Understanding the plastic deformation mechanisms and behaviour are thus crucial but challenging in the Ti-Al system. Experimental studies have revealed a rich set of unusual deformation phenomena associated with dislocation core structures and energetics. For example, L1\(_\text{0}\)-TiAl single crystals exhibit anomalous yield behaviour at 700-1000 \(^{\circ}\)C~\cite{inui_1997_pma}, while D0\(_{19}\)-\ce{Ti3Al} single crystals have strong plastic anisotropy~\cite{inui_1993_pma} with both critical resolved shear stresses (CRSS) and tensile elongation differing by more than an order of magnitude among the different slip systems~\cite{inui_1993_pma}. Transmission electron microscopy (TEM) studies show that dislocations in TiAl tend to align in the screw orientation with frequent pinning~\cite{hug_1988_pma,sriram_1997_pma,zghal_1998_am,couret_1999_pma,katzarov_2011_am}, while the \ccadisl superdislocations decompose into non-planar, climb-dissociated core structures~\cite{wiezorek_2010_mmta}. These unusual defect structures appear to be in locked configurations and thus can strongly influence plastic flow, hardening and fracture behaviour. However, their origins (intrinsic or extrinsic), mechanisms of formation, and dependences on temperature and alloy compositions are not well understood.

Given the importance of the Ti-Al system, extensive modelling works have been carried out to shed light on lattice defect structures and energetics at atomistic resolutions~\cite{appel_2016_pms}.  Density functional theory (DFT) calculations have been performed to determine the relevant generalized stacking fault energies (GSFEs), including the superlattice intrinsic stacking fault (SISF~\cite{fu_1990_pml,liu_2007_im,sun_2013_sm,kanani_2016_am,jeong_2018_sr}), anti-phase boundary (APB~\cite{fu_1990_pml,kanani_2016_am,jeong_2018_sr}), complex stacking fault (CSF~\cite{sun_2013_sm,kanani_2016_am,jeong_2018_sr}) on the \(\{111\}\) slip plane of \(\gamma\)-TiAl, as well as the SISF~\cite{liu_2007_im}, APB~\cite{koizumi_2006_pm} on the \(\{0001\}\)-plane, APB~\cite{koizumi_2006_pm} on the \(\{1\bar{1}00\}\) slip plane of \(\alpha_2\)-\ce{Ti3Al}.  However, substantial discrepancies exist between these first-principles based GSFEs and values extrapolated from experiments based on partial dislocation separation distances~\cite{karkina_2012_msmse}.  For example, on the prism \((1\bar{1}00)_\text{I}\) and \((1\bar{1}00)_\text{II}\) planes in \ce{Ti3Al}, DFT-based APB energies are more than 2 times the experimental values~\cite{legros_1996_pma,koizumi_2006_pm,karkina_2012_msmse}.  Moreover, DFT calculations are computationally expensive and limited to small supercell sizes of a few thousand valence electrons in routine total energy/structure optimization calculations.  They are also often performed at 0~K without consideration of entropic effects on defect properties.  First-principles DFT-based molecular dynamics (MD) simulations are even more challenging.  They are currently not practical for direct determination of superdislocation core structures and their temperature-dependent glide behaviour in the Ti-Al system.  More fundamentally, even the GSFEs on \ce{Ti3Al} pyramidal planes have not been quantitatively determined at DFT accuracies to the best of our knowledge.

The length-scale and time-scale limitations of DFT calculations are well recognized across the materials modelling community.  To overcome these limitations, many classical interatomic potentials (IAPs) have been developed specifically for the TiAl and \ce{Ti3Al} intermetallic alloys using the embedded-atom method (EAM~\cite{daw_1984_prb,farkas_1996_msmse,zope_2003_prb}), the modified embedded-atom method (MEAM~\cite{baskes_1987_prl,baskes_1989_prb,kim_2016_cms}) and the bond order potential~\cite{znam_2003_pm}.  Recently, machine learning (ML) IAPs have also been developed using polynominal invariants to represent neighbouring atomic density (e.g., MLP3 by Seko~\cite{seko_2020_prb}).  However, these IAPs have limited capabilities in reproducing lattice defect properties in both phases~\cite{pei_2021_cms}.  In particular, the classical IAPs~\cite{farkas_1996_msmse,kim_2016_cms} based on the EAM and MEAM have considerable deficiencies in reproducing the elastic constants and SF energies of the L1\(_0\)-TiAl and D0\(_{19}\)-\ce{Ti3Al} phases (see Ref.~\cite{pei_2021_cms}), while the BOP~\cite{znam_2003_pm} is fit to TiAl and is less accurate for the \ce{Ti3Al} phase (e.g., \(C^{\text{BOP}}_{44} = 45\) GPa vs. \(C^{\text{EXP}}_{44} = 64\) GPa~\cite{znam_2003_pm}).  Specifically, the EAM potentials~\cite{farkas_1996_msmse,zope_2003_prb} and the MEAM potential~\cite{kim_2016_cms} do not reproduce the negative Cauchy pressures (\(C_{13}-C_{44}\) and \(C_{12} - C_{66}\)) as measured from experiments and have unusual, negative SF energies on pyramidal planes of the D0\(_{19}\)-\ce{Ti3Al} structure.  The MLP3 shows some promises in modelling the Ti-Al system.  However, this potential significantly underestimates the surface energies in both structures (e.g., \(\gamma^{\text{MLP3}}_{\text{L1}_0-\{111\}} = 0.874 \) J/m\(^{2}\) vs. \(\gamma^{\text{DFT}}_{\text{L1}_0-\{111\}}= 1.667 \) J/m\(^{2}\)) and is unsuitable for modelling fracture processes in TiAl alloys where brittle cleavage is commonly observed~\cite{ding_2015_im}.  Therefore, no accurate IAPs exist at present for general modellings of lattice and interfacial defects in the technologically important Ti-Al material system.

In this work, we develop a new ML-IAP suitable for modelling all dislocation core structures and cleavage properties in the L1\(_\text{0}\)-TiAl and D0\(_{19}\)-\ce{Ti3Al} intermetallic phases. We employ the moment tensor potential (MTP~\cite{shapeev_2016_mms}) framework. Specifically, the MTP is trained using accurate DFT energies and forces of a comprehensive set of structural configurations, comprising ground-state structures, \textit{ab initio} MD snapshots, surface structures, solid-solution structures and strained bulk structures. The resulting MTP exhibits superior accuracy when compared to all previous IAPs. It reproduces the formation energies/phase diagram within a composition range of Ti-(25-66)at.\%Al where commercial TiAl intermetallic alloys are based. Furthermore, we compute the relevant \(\gamma\)-lines and \(\gamma\)-surfaces in the two phases using DFT for the first time and show that the MTP accurately reproduces nearly all fundamental (i) lattice and elastic properties and their temperature-dependence; (ii) surface energies; and (iii) \(\gamma\)-lines in both the L1\(_0\)-TiAl and D0\(_{19}\)-\ce{Ti3Al} structures. Using the new MTP, we calculate all relevant dislocation core structures and compare them with extant DFT/BOP/experimental results. In the L1\(_0\) structure, all screw and edge dislocations adopt near-planar dissociations upon structure optimization. In the D0\(_{19}\) structure, the screw \adisl and \ccadisl dislocations have non-planar dissociations on basal and pyramidal planes, respectively. The mixed and edge \ccadisl are not stable on the pyramidal planes and undergo a pyramidal to basal climb-dissociation, largely consistent with previous TEM observations in elemental HCP metals. The new MTP opens the path to realistic modelling of dislocation and fracture behaviour in \(\gamma\)-TiAl and \(\alpha_2\)-\ce{Ti3Al}.

In the following, we first briefly describe the procedure and datasets for training the MTP, followed by details of DFT setup in both the MTP training and lattice properties calculations. Section~\ref{sec:result} presents the properties of the optimized MTP in comparison with DFT and other available data. We benchmark and analyze a comprehensive set of bulk lattice and defect properties, including lattice and elastic constants, surface energies, \(\gamma\)-surfaces and \(\gamma\)-lines, as well as all dislocation core structures and dissociations. Section~\ref{sec:disc} discusses the overall performance of the MTP and its predicted dislocation cores with respect to previous results, followed by a brief conclusion of the current work in Section~\ref{sec:conc}.

\section{Computational Methods and Simulation Models}

\subsection{Moment tensor potential}
The MTP formalism has been extensively discussed in earlier works~\cite{shapeev_2016_mms,gubaev_2019_cms,zuo_2020_jpca} and successfully applied to a wide range of systems and problems, including elemental metals~\cite{shapeev_2016_mms,novoselov_2019_cms,zuo_2020_jpca}, boron~\cite{podryabinkin_2019_prb}, alloys~\cite{gubaev_2019_cms}, gas-phase reactions~\cite{novikov_2018_pccp}, cathode coating materials~\cite{wang_2020_cm} and Li superionic conductors~\cite{qi_2021_mtp}. Briefly, the MTP describes the local environment around each atom in terms of moment tensors $M_{\mu,\nu}$, defined as
\begin{equation}
  M_{\mu,\nu}(\mathbf{n_i}) = \sum_{j}f_{\mu}(|\mathbf{r_{ij}}|,z_i,z_j)\underbrace{\mathbf{r_{ij}} \otimes... \otimes \mathbf{r_{ij}}}_\textrm{$\nu$ times}. \label{eqn:mtp_descriptor}
\end{equation}
Here, $\mathbf{n_i}$ denotes the $i^\text{th}$ atom type, relative position, and all of its neighboring atoms. $z_i$ and $z_j$ represent the atomic types of the $i^\text{th}$ atom and its $j^\text{th}$ neighbor, respectively, and $\mathbf{r_{ij}}$ is the position vector of the $j^\text{th}$ neighbor to the $i^\text{th}$ atom. \(z_{i/j}\) is an integer from 0 to \(n-1\) for a system with \(n\) different types of atoms. The radial part of the atomic environment is given by the $f_{\mu}$ term, and the angular part is encoded by the outer product ($\otimes$) of the $\mathbf{r_{ij}}$ vectors, which is a tensor of rank $\nu$. The MTP then contracts the moment tensors $M_{\mu,\nu}$ to basis functions and applies regression to relate energies, forces and stresses of a material system to the basis functions.  In this work, the energy, force and stress data points are assigned weights of 100:1:0, similar to previous works~\cite{li_2018_prb,deng_2019_npjcm,zuo_2020_jpca,li_2020_npjcm,qi_2021_mtp}.

In MTPs, two key parameters control the accuracy-efficiency tradeoff. The cutoff radius $r_\text{c}$ determines the maximum interaction range between atoms, while the maximum level $lev_\text{max}$, an even integer ranging from 2 to 28, controls the number of parameters in the moment tensors $M_{\mu,\nu}$ and thus the model complexity. In this work, a grid search for an optimal MTP was performed with $r_\text{c}$ ranging from 4.4 to 7 \r{A}\ with 0.2 \r{A}\ intervals, and $lev_\text{max}$ from 18 to 24, respectively. For each combination of $r_\text{c}$ and $lev_\text{max}$, five MTPs were created with random initial states, for a total of 280 MTPs trained. Among all the MTPs, the MTP with $r_\text{c} = 4.8$ \r{A}\ and $lev_\text{max} = 22$ had the lowest mean absolute error (MAE) of elastic constants of the 6 stable structures at stoichiometic compositions in the Ti-Al phase diagram.

All training, evaluations and simulations with MTP were performed using MLIP~\cite{shapeev_2016_mms,gubaev_2019_cms}, LAMMPS~\cite{thompson_2022_cpc} and the open-source Materials Machine Learning (maml) Python package~\cite{chen_maml_2020}.

\subsection{Training structure generation}
A robust and diverse set of training structures is essential to developing accurate MTPs. Based on our previous experience in the Ni-Mo system~\cite{li_2018_prb}, we used 5 categories of training structures:
\begin{enumerate}
  \item 33 ground-state polymorph structures in the Ti-Al system stored in the Materials Project (MP~\cite{jain_2013_aplm}) database.
  \item 2160 configurations from NVT \textit{ab initio} molecular dynamics (AIMD) simulations of the 6 stable structures at stoichiometic compositions in the Ti-Al phase diagram in MP, i.e., Ti (P6/mmm, mp-72), \(\upalpha_2\)-\ce{Ti3Al} (P6$_3$/mmc, mp-1823), $\upgamma$-TiAl (P4/mmm, mp-1953), \ce{TiAl2} (I4$_1$/amd, mp-567705), \ce{TiAl3} (I4/mmm, mp-542915) and Al (Fm-3m, mp-134). The AIMD simulations are performed at 300, 1000 and 3000 K with 90\%, 100\% and 110\% of the respective ground-state volumes. 40 configurations were extracted from each AIMD run lasting 10 ps.
  \item 185 surface structures with Miller indices up to three from the 6 stable phases at stoichiometic compositions in the Ti-Al phase diagram in MP and the HCP Ti (P6$_3$/mmc, mp-46).
  \item 580 solid-solution alloy structures constructed by partial substitution of Al with Ti in supercells of the bulk fcc Al (mp-134) and partial substitution of Ti with Al in the two Ti phases, i.e., bulk hexagonal (P6/mmm, mp-72) and HCP (P6$_3$/mmc, mp-46). Compositions of the solid solution Al$_{x}$Ti\(_{1 - x}\) were generated with \(x\) ranging from 12.5 to 87.5 at.\% at intervals of 12.5 at.\%. For each substitution scenario, up to 30 structures were selected from the enumeration.
  \item 840 homogeneously deformed structures constructed by applying strains within $\pm$10\% at 1\% intervals and in 6 different modes (\(\varepsilon_\text{xx},\varepsilon_\text{yy},\varepsilon_\text{zz},\varepsilon_\text{xy},\varepsilon_\text{yz},\varepsilon_\text{yz}\)) to bulk supercells of 6 stable phases at stoichiometic compositions in the Ti-Al phase diagram in MP. To correct the erroneous low energy of cubic TiAl$_3$ (Pm-3m, mp-998981) predicted by the potential, two polymorphs of TiAl$_3$ are included.
\end{enumerate}

Overall, a 90:10 train:test split was selected to the above data set.

\subsection{DFT calculations for MTP training}
\label{sec:dft_mtp_train}

DFT calculations were performed using the Vienna \textit{ab initio} simulation package (VASP~\cite{kresse_1996_cms,kresse_1996_prb}).  The exchange-correlation functional used was the Perdew-Burke-Ernzerhof (PBE~\cite{perdew_1996_prl}) generalized gradient approximation (GGA). The outer shell $3p^63d^34s^1$ and $3s^23p^1$ electrons are treated as valence electrons for Ti and Al, respectively.  Core electrons are replaced using the projector augmented-wave (PAW~\cite{blochl_1994_prb}) pseudopotentials.  For structural optimization, spin-polarized DFT is employed with an energy cutoff of 520 eV and a \(k\)-point density of at least 100/\r{A}$^{3}$, consistent to the default setting in the MP~\cite{jain_2013_aplm}.  The AIMD simulations are performed with a single $\Gamma$ \(k\)-point and nonspin-polarized DFT.  To keep consistency in the training energies and forces,  DFT self-consistent calculations are performed on the snapshots using the same parameters as the rest of the data.

\subsection{DFT Calculation of lattice defect properties}
\label{sec:dft_defect}

The GSFE \(\gamma\)-surfaces are calculated using the vacuum-slab method (15 \AA\ vacuum spacing) and the \(\gamma\)-lines are calculated using the tilt-cell method~\cite{nieminen_1992_am,sutton_1995_interfaces,kibey_2006_apl,yin_2017_am}.   While these calculations were performed previously and DFT parameters are different from that used in the MTP training above, they yield nearly identical or similar results in both bulk and defect properties; the calculations here are mainly for benchmarking/validation of various potentials and thus do not affect the general conclusion of the current work.  In DFT, core electrons are replaced by PAW pseudopotentials with 10 ($3p^63d^34s^1$) and 3 ($3s^23p^1$) valence electrons for Ti and Al~\cite{blochl_1994_prb,kresse_1996_prb}. The exchange-correlation is treated using the PW91 GGA functional~\cite{perdew_1992_prb}.  We employed a kinetic energy cutoff of 520 eV for the plane-wave basis set, and the first-order Methfesel-Paxton smearing~\cite{methfessel_1989_prb} with a smearing width of 0.2 eV and \(\Gamma\)-centered Monkhorst-Pack \(k\)-point meshes~\cite{monkhorst_1976_prb} for the Brillion-zone integration. The in-plane \(k\)-point grid densities are equivalent to at least \(12\times12\times12\) \(k\)-points in the L1\(_0\) unit cell and \(10 \times 10 \times 12\) \(k\)-points in the D0\(_{19}\) unit cell and 1 \(k\)-point is used in the out-of-plane direction.

For the surface energy calculations, a vacuum spacing of 15 \AA\ is used. The structure is optimized with all supercell vectors fixed and all atoms allowed to undergo ionic relaxation. For the \(\gamma\)-surface, the supercell is constructed with the slip plane in the \(x-y\) plane and the slip plane normal in the \(z\)-direction. The \(\gamma\)-surfaces are calculated with ion positions optimized in the slip-plane-normal direction while the \(\gamma\)-lines are calculated using the tilted supercells with further ionic optimization in the two directions perpendicular to the slip direction and supercell vectors optimized in the slip plane normal to achieve zero normal stresses. The meta-stable SFs are calculated with all ionic constraints removed. Convergence is assumed when all ionic forces drop below 10 meV/\AA.

Other lattice properties and dislocation core structures are calculated using standard methods in LAMMPS. All atomistic structures are visualized using the Open Visualization Tool (OVITO~\cite{stukowski_2009_msmse}) and dislocation cores are illustrated using the differential displacement (DD) map~\cite{vitek_1970_pma}. The details are described in a recent work~\cite{wang_2022_prm}. The potential is available at \href{https://github.com/materialsvirtuallab/maml/tree/master/mvl_models/pes/TiAl/mtp.2023.04}{maml} and can be directly used with LAMMPS and MLIP by interested readers.

\section{Results}
\label{sec:result}

\subsection{Ti-Al formation energies}

\begin{figure*}[!htbp]
 \centering
 \includegraphics[width=0.70\textwidth]{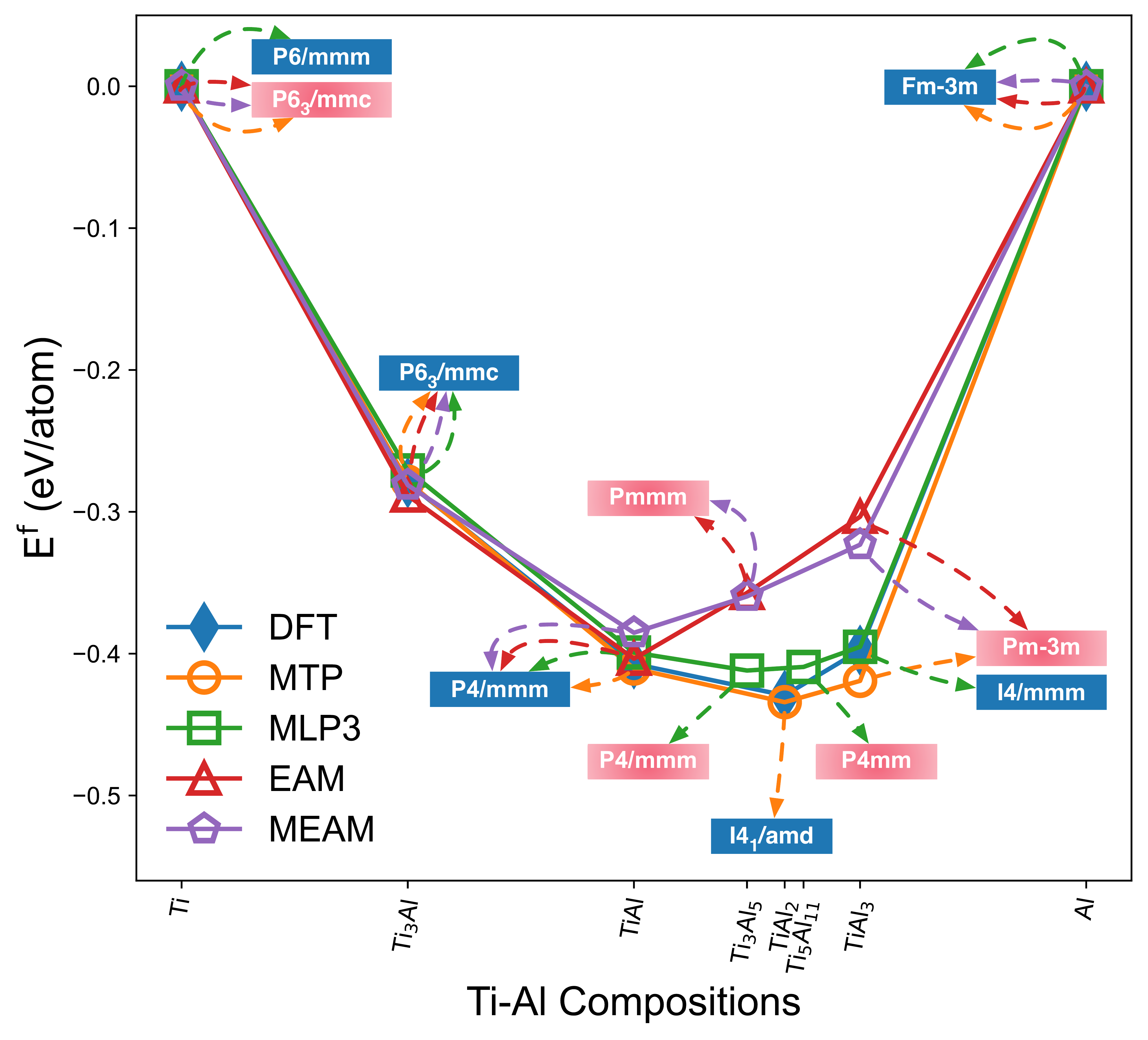}
 \caption{Formation energies \(\Delta E^\text{P}_{mn}\) of stable phases in the Ti-Al binary system calculated by interatomic potentials (EAM~\cite{zope_2003_prb}, MEAM~\cite{kim_2016_cms}, MLP3~\cite{seko_2020_prb} and MTP) and DFT. The stable structures predicted by DFT are shown in blue color; the stable structures predicted by interatomic potentials are shown in red color if they are different.  Within the composition range of interest (25at.\%Al- 66at.\%Al), only the MTP reproduces all the stable phases in agreement with DFT. See Ref.~\cite{mehl_2017_cms} for the space group nomenclatures. } 
 \label{fig:form_energy}
\end{figure*}

Figure~\ref{fig:form_energy} shows the formation energies \(\Delta E^{\text{P}}_{mn}\) of stable phases in the binary Ti-Al system. Here, the formation energy is defined as
\begin{align}
 \Delta E^{\text{P}}_{mn} = \dfrac{E^{\text{P}}_{mn} - mE_\text{Ti} - nE_\text{Al}}{m+n},
\end{align}
where \(m\) and \(n\) are the numbers of Ti and Al atoms in the primitive cell of phase \(\text{P}\), \(E^{\text{P}}_{mn} \) is the energy per chemical formula of phase \(\text{P}\), and \(E_\text{Ti}\) and \(E_\text{Al}\) are the energies per atom of Ti and Al in the HCP and FCC reference structures, respectively. We focus on the \(\upalpha_2\)-\ce{Ti3Al}, \(\upgamma\)-TiAl, \ce{TiAl2}, and \(\upalpha\)-\ce{TiAl3} phases. The classical EAM~\cite{zope_2003_prb} and MEAM~\cite{kim_2016_cms} IAPs have large discrepancies with respect to DFT in the \ce{TiAl2} and \(\upalpha\)-\ce{TiAl3} phases, while the ML-IAPs have better overall matches. Only the MTP has successfully identified a stable \ce{TiAl2} phase in agreement with DFT, while the \ce{TiAl2} phase is unstable in the other 3 IAPs. On the other hand, only the MLP3 correctly identifies the most stable structure for pure Ti (P6/mmm-\(\omega\)) and \ce{TiAl3} (I4/mmm-\(\alpha\)) phases. This is not surprising since the relative energy difference between the polymorph structures of these two compositions to the respective convex hull are small according to DFT, i.e. 4 and 25 meV/atom for \ce{Ti} and \ce{TiAl3}, respectively. MLP3 and MTP are in general more accurate, as they are trained with a broad range of structure prototypes. More importantly, only the MTP reproduces all the stable phases within the composition range of interest, i.e., 25at.\%Al- 66at.\%Al, which make it suitable for atomistic modelling in the L1\(_0\) \(\upgamma\)-TiAl and D0\(_{19}\) \(\upalpha_2\)-\ce{Ti3Al} structures.

\begin{figure*}[!htbp]
 \centering
 \includegraphics[width=0.85\textwidth]{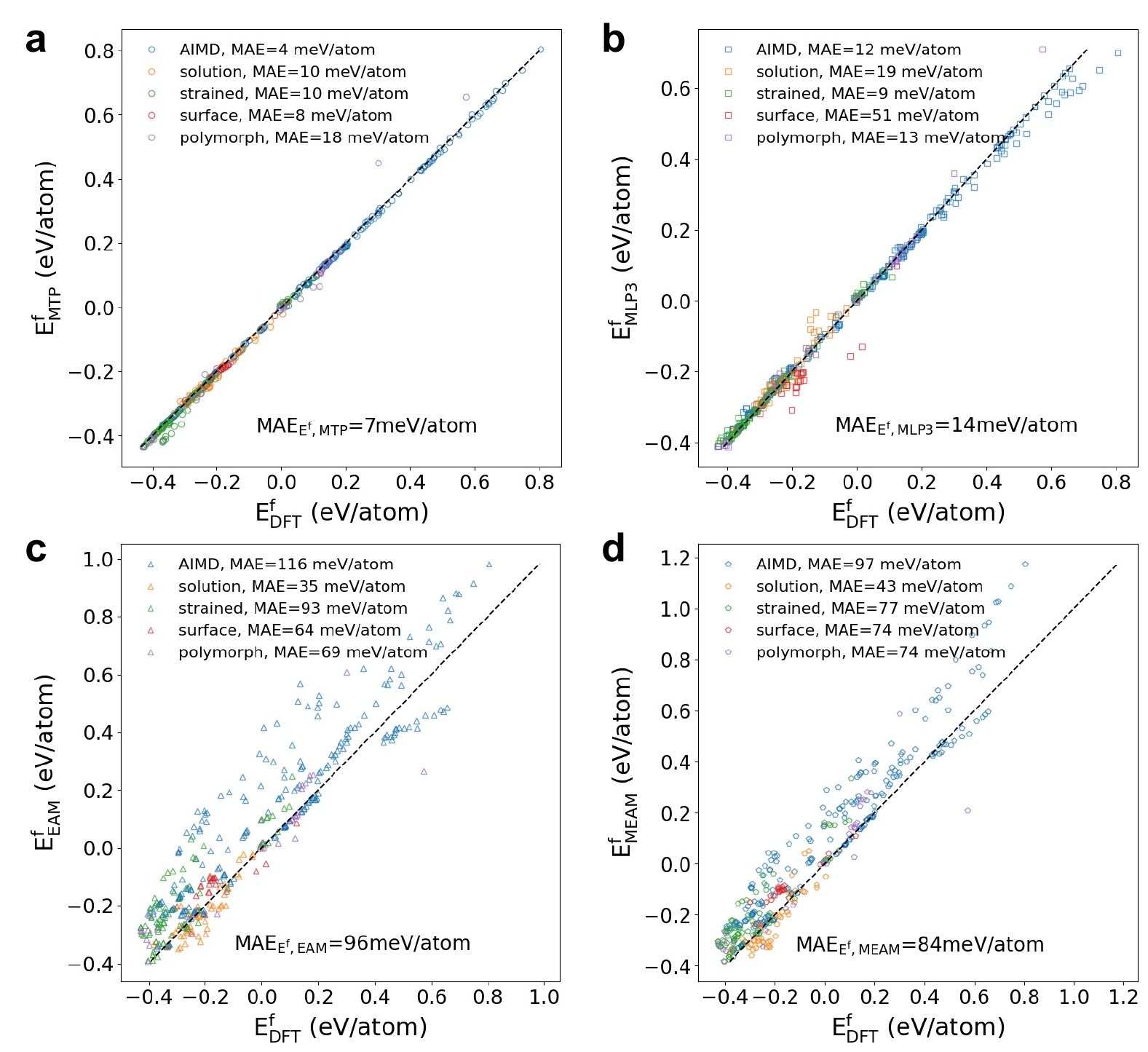}
 \caption{\label{fig:delta_E_random} Distribution of error in formation energies of 5 groups of structures predicted by interatomic potentials with respect to DFT. (a) MTP. (b) MLP3~\cite{seko_2020_prb}. (c) EAM~\cite{zope_2003_prb}. (d) MEAM~\cite{kim_2016_cms}. The MTP has the smallest overall mean absolute error (MAE). }
\end{figure*}

Figure~\ref{fig:delta_E_random} shows the distributions of discrepancies with respect to DFT in the formation energies of 5 groups of structures.  For all the cases, the two ML-IAPs have overall MAEs nearly an order of magnitude smaller than the two classical IAPs. Within the two ML-IAPs, the MLP3 has significantly larger MAE for surface structures than the MTP (51 meV/atom vs. 8 meV/atom), while the MTP has a slightly larger MAE in the polymorph structures (18 meV/atom vs. 13 meV/atom). Overall, both ML-IAPs have better reproducibility in the formation energies of these structures, indicating their better transferability in modelling complex structures of varying local atomic environments.

\subsection{Equation of states}
\begin{figure*}[!htbp]
 \centering
 \includegraphics[width=0.70\textwidth]{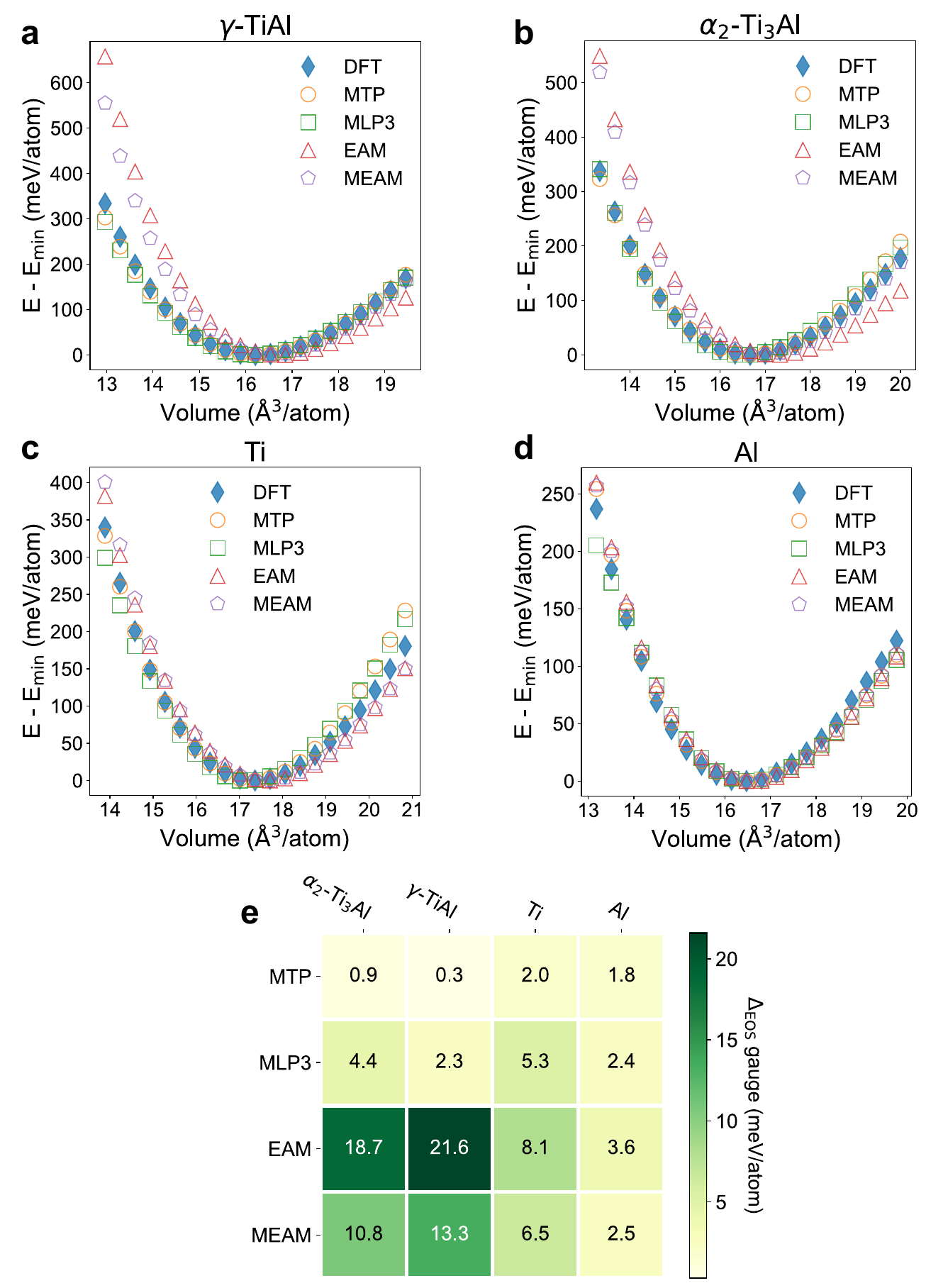}
 \caption{\label{fig:eos_comp} Equation of state predicted by 4 interatomic potentials and DFT. (a) L1\(_0\) \(\gamma\)-TiAl. (b) D0\(_{19}\) \(\alpha_2\)-\ce{Ti3Al}. (c) HCP Ti (P6$_3$/mmc). (d) FCC Al (Fm$\bar{3}$m). (e) Comparisons of \(\Delta_\text{EOS}\) (Eq.~\eqref{eq:delta_eos}) among the potentials (MTP, MLP3~\cite{seko_2020_prb}, EAM~\cite{zope_2003_prb} and MEAM~\cite{kim_2016_cms}) with respect to DFT prediction.} 
\end{figure*}

Figure~\ref{fig:eos_comp} shows the equation of state (EOS) of 4 structures (L1\(_0\)-TiAl, D0\(_{19}\)-\ce{Ti3Al}, HCP-Ti and FCC-Al) predicted by the IAPs and DFT. Here, the accuracy is measured by $\Delta_\text{EOS}$~\cite{lejaeghere_2013_crssms} defined as
\begin{align}\label{eq:delta_eos}
 \Delta_\text{EOS} = \sqrt{\frac{\int{^{1.06V_o}_{0.94V_o}[E_a(V)-E_b(V)]^2 \text{d}V}}{0.12V_o}},
\end{align}
where $E_a(V)$ and $E_b(V)$ are energies per atom at volume $V$ computed using models \(a\) and \(b\) (i.e., IAPs and DFT), respectively, and \(V_o\) is the equilibrium volume in DFT. The two classical IAPs are developed based upon the potentials for pure Ti and Al, so their accuracies are relatively better in the elemental structures than in the intermetallic structures. In particular, they exhibit relatively large discrepancies, in the range of 10-22 meV/atom, with respect to DFT in the L1\(_0\)-TiAl and D0\(_{19}\)-\ce{Ti3Al} structures. On the other hand, the ML-IAPs are trained based on configurations of multiple solid solution and intermetallic structures and thus have smaller deviations from the respective DFT values. For all four structures, the MTP has \(\Delta E_\text{EOS} < \) 2 meV/atom, comparable to the $\Delta E_\text{EOS}$ between different pseudopotential implementations of DFT relative to all-electron calculations~\cite{lejaeghere_2013_crssms}.  The MTP is thus promising for simulations of bulk properties under a wide range of volumetric strains and pressures.

\subsection{Lattice constants and elastic constants}

Table~\ref{tab:lat_prop_0K} shows the 0K lattice and elastic constants predicted by the IAPs and DFT in comparison with experimental values.  Some discrepancies also exist between DFT and experiments, e.g., \(C^\text{DFT}_{44} = 42 \ \text{GPa}\) vs. \(C^\text{EXP}_{44} = 51 \ \text{GPa}\) in HCP Ti.  The benchmark can thus be carried out with respect to DFT, since ML potentials are fit to datasets from DFT.  Nonetheless, the general conclusion does not depend on this choice of references.  Overall, both ML-IAPs have accurate elastic properties in the intermetallic structures while the classical IAPs have better accuracies in elemental HCP Ti and FCC Al.  For the MLP3, the largest discrepancies are in HCP Ti (35\% and 31\% in \(C_{13}\) and \(C_{12}\)), FCC Al (32\% and 22\% in \(C_{11}\) and \(C_{44}\)) and L1\(_0\)-TiAl (\(-\)22\%, 20\% and 17\% in \(C_{66}\), \(C_{13}\) and \(C_{12}\)).  The MTP has similar discrepancies in HCP Ti as the MLP3; both overestimate \(C_{12}\) and \(C_{13}\) by more than 15\%.  The MTP also underestimates \(C_{44}\) of HCP Ti by 33\%.  In addition, its \(C_{12}\) of the L1\(_0\) and D0\(_{19}\) structures are 23\% larger than the experimental values, while the rests are within 15\% from the corresponding experimental values.   On the other hand, the two classical potentials (EAM~\cite{zope_2003_prb} and MEAM~\cite{kim_2016_cms}) have larger discrepancies exceeding 40\% in the two intermetallic phases.

\begin{table*}[!htbp]
 \centering
 \caption{Lattice constants \(a_0\) and \(c_0\) (\r{A}) and elastic constants \(C_{ij}\) (GPa) of FCC Al, HCP Ti, L1\(_{0}\) \(\gamma\)-TiAl and D0\(_{19}\) \(\alpha_2\)-\ce{Ti3Al} calculated by 4 potentials and DFT at 0 K. The experimental elastic constants of Al and Ti are based on measurements at 4~K~\cite{simmons_1971_mit}.  The elastic constants of TiAl are measurements of a \(\gamma\)-Ti-56at.\%Al alloy extrapolated to 0~K~\cite{he_1997_msea}.  The percentage errors with respect to experimental values are shown in the parenthesis and errors exceeding \(\pm\)15\% are highlighted in bold font.
 }
 \small
  \begin{tabular}{llllllll}
   \toprule
  \textbf{Structure} & \textbf{Property} & \textbf{Experiment} & \textbf{DFT} & \textbf{MTP} & \textbf{MLP3}~\cite{seko_2020_prb} & \textbf{EAM}~\cite{zope_2003_prb} & \textbf{MEAM}~\cite{kim_2016_cms} \\
\hline
\multirow{10}{*}{L1\(_{0}\) TiAl} & \(a_0\) & 3.988~\cite{he_1997_msea} & 3.996 (0.2\%)~\cite{zhang_2017_ijmpb} & 3.993 (0.1\%) & 3.977 (\(-\)0.3\%) & 3.998 (0.2\%) & 4.018 (0.8\%) \\
& \(c_0\) & 4.067~\cite{he_1997_msea} & 4.076 (0.2\%)~\cite{zhang_2017_ijmpb} & 4.075 (0.2\%) & 4.080 (0.3\%) & 4.187 (2.9\%) & 4.099(0.8\%) \\
& \(C_{11}\) & 187~\cite{he_1997_msea} & 164 (\(-\)12\%)~\cite{fu_2010_im} & 175 (\(-\)6\%) & 167 (\(-\)11\%) & 197 (5\%) & 181 (\(-\)3\%) \\
& \(C_{12}\) & 75~\cite{he_1997_msea} & 86 (\textbf{15}\%)~\cite{fu_2010_im} & 92 (\textbf{23}\%) & 88 (\textbf{17}\%) & 107 (\textbf{43\%}) & 76 (1\%) \\
& \(C_{13}\) & 75~\cite{he_1997_msea} & 81 (8\%)~\cite{fu_2010_im} & 82 (9\%) & 90 (\textbf{20}\%) & 114 (\textbf{52\%}) & 134 (\textbf{79\%}) \\
& \(C_{33}\) & 183~\cite{he_1997_msea} & 179 (\(-\)2\%)~\cite{fu_2010_im} & 165 (\(-\)10\%) & 176 (\(-\)4\%) & 213 (\textbf{16\%}) & 234 (\textbf{28\%}) \\
& \(C_{44}\) & 109~\cite{he_1997_msea} & 110 (1\%)~\cite{fu_2010_im} & 114 (5\%) & 114 (5\%) & 92 (\(-\)\textbf{16}\%) & 86 (\(-\)\textbf{21\%}) \\
& \(C_{66}\) & 81~\cite{he_1997_msea} & 73 (\(-\)10\%)~\cite{fu_2010_im} & 70 (\(-\)14\%) & 63 (\(-\)\textbf{22\%}) & 85 (5\%) & 62 (\(-\)\textbf{23\%}) \\
& \(C_{12}-C_{66}\) & \(-\)6 & 13 & 22 & 25 & 22 & 14\\
& \(C_{13}-C_{44}\) & \(-\)34 & \(-\)29 & \(-\)32 & \(-\)24 & 22 & 48 \\
\\
\multirow{10}{*}{D0\(_{19}\) \ce{Ti3Al}} & \(a_0\) &5.77~\cite{pearson_1958_hb} & 5.76 (\(-0.2\%\))~\cite{zhang_2017_ijmpb} & 5.762 (\(-\)0.1\%) & 5.729 (\(-\)0.7\%) & 5.784 (0.2\%) & 5.805 (0.6\%) \\
& \(c_0\) & 4.62~\cite{pearson_1958_hb} & 4.66 (0.9\%)~\cite{zhang_2017_ijmpb} & 4.644 (0.5\%) & 4.644 (0.5\%) & 4.750 (2.8\%) & 4.655 (\(-\)0.7\%) \\
& \(C_{11}\) & 183~\cite{tanaka_1996_pma} & 192 (5\%)~\cite{zhang_2017_ijmpb} & 193 (6\%) & 196 (7\%) & 199 (9\%) & 201 (10\%) \\
& \(C_{12}\) & 89~\cite{tanaka_1996_pma} & 81 (\(-\)9\%)~\cite{zhang_2017_ijmpb} & {110} (\textbf{23\%}) & 99 (11\%) & 89 (0\%) & 108 (\textbf{21\%}) \\
& \(C_{13}\) & 63~\cite{tanaka_1996_pma} & 63 (0\%)~\cite{zhang_2017_ijmpb} & 70 (12\%) & 71 (13\%) & 74 (\textbf{17\%}) & 91 (\textbf{45\%}) \\
& \(C_{33}\) & 225~\cite{tanaka_1996_pma}& 233(4\%)~\cite{zhang_2017_ijmpb} & 236 (5\%) & 231 (3\%) & 224 (0\%) & 239 (6\%) \\
& \(C_{44}\) & 64~\cite{tanaka_1996_pma} & 62 (\(-\)3\%)~\cite{zhang_2017_ijmpb} & 59 (\(-\)8\%) & 63 (\(-\)2\%)     & 51 (\(-\)\textbf{20\%}) & 46 (\(-\)\textbf{29\%}) \\
& \(C_{66}\) & 47~\cite{tanaka_1996_pma} & 56 (\textbf{19\%})~\cite{zhang_2017_ijmpb} & 42 (\(-\)11\%) & 48 (2\%) & 55 (\textbf{17\%}) & 46 (\(-\)1\%) \\
& \(C_{12}-C_{66}\) & 42 & 25& 68 & 51 & 34 & 62 \\
& \(C_{13}-C_{44}\) & \(-\)1 & 1& 11 & 8 & 23 & 46 \\
   \\
\multirow{4}{*}{FCC Al} & \(a_0\)  & 4.046~\cite{pham_2011_prb} & 4.039 (\(-\)0.2\%)~\cite{jain_2013_aplm} & 4.042 (\(-\)0.1\%) & 4.041 (\(-\)0.1\%) & 4.050 (0.1\%) & 4.045 (0\%) \\
& \(C_{11}\) & 107~\cite{simmons_1971_mit} & 111 (4\%)~\cite{sinko_2002_jpcm} & 116 (8\%) & 141 (\textbf{32\%}) & 120 (12\%) & 120 (12\%) \\
& \(C_{12}\) & 61~\cite{simmons_1971_mit} & 58 (\(-\)5\%)~\cite{sinko_2002_jpcm} & 59 (\(-\)4\%) & 61 (\(-\)1\%) & 58 (\(-\)5\%) & 58 (\(-\)5\%) \\
& \(C_{44}\) & 28~\cite{simmons_1971_mit} & 31 (11\%)~\cite{sinko_2002_jpcm} & 28 (0\%) & 34 (\textbf{22\%}) & 29(\(-\)5\%) & 28(0\%) \\
\\
\multirow{7}{*}{HCP Ti} & \(a_0\) & 2.947~\cite{boyer_2005_jmep} & 2.934 (\(-\)0.4\%)~\cite{jain_2013_aplm} & 2.945 (\(-\)0.1\%) & 2.949 (0.2\%) & 2.953 (0.2\%) & 2.954 (\(-\)0.1\%) \\
& \(c_0\) & 4.674~\cite{boyer_2005_jmep} &4.657 (\(-\)0.4\%)~\cite{jain_2013_aplm} & 4.605 (\(-\)1.5\%) & 4.567 (\(-\)2.3\%) & 4.681 (0.1\%) & 4.687 (0.3\%) \\
& \(C_{11}\) &176~\cite{simmons_1971_mit} &177 (1\%)~\cite{jain_2013_aplm} & 186 (6\%) & 175 (\(-\)1\%) & 171 (\(-\)3\%) & 170 (\(-\)3\%) \\
& \(C_{12}\) &87~\cite{simmons_1971_mit} &83 (\(-\)5\%)~\cite{jain_2013_aplm} & 116 (\textbf{33\%}) & 114 (\textbf{31\%}) & 84 (\(-\)3\%) & 80 (\(-\)8\%) \\
& \(C_{13}\) &68~\cite{simmons_1971_mit} &76 (12\%)~\cite{jain_2013_aplm} & 82 (\textbf{21\%}) & 92 (\textbf{35\%}) & 77 (13\%) & 75 (10\%) \\
& \(C_{33}\) &191~\cite{simmons_1971_mit} &191 (0\%)~\cite{jain_2013_aplm} & 182 (\(-\)5\%) & 222 (\textbf{16\%}) & 190 (\(-\)1\%) & 187 (\(-\)2\%) \\
& \(C_{44}\) &51~\cite{simmons_1971_mit} &42 (\(-\)\textbf{18\%})~\cite{jain_2013_aplm} & 34 (\(-\)\textbf{33\%}) & 58 (14\%) & 53 (4\%) & 42 (\(-\)\textbf{18\%}) \\
   \bottomrule
  \end{tabular}
 \label{tab:lat_prop_0K}
\end{table*}

In addition, the MTP reproduces the negative Cauchy pressure \(C_{13} - C_{44} = -32 \) GPa of the L1\(_0\)-TiAl structure, which is close to the experimental value of \(-\)34 GPa~\cite{tanaka_1996_pml,he_1997_msea}. The MLP3 also exhibits negative \(C_{13} - C_{44} = -24 \) GPa. The negative Cauchy pressure arises from the valence \(sp\) electrons and has been challenging to reproduce by classical IAPs~\cite{znam_2003_pm}. The ML-IAPs seem to have intrinsic advantages in reproducing such complex properties. On the other hand, both the MTP and MLP3 exhibit relatively large discrepancies in the elastic properties of unary HCP Ti and FCC Al. This is not surprising since these ML-IAPs are trained extensively on datasets of the intermetallic structures. In addition, it also suggests that accuracies in particular properties or phases are not always guaranteed in ML-IAPs if they are not specifically trained.

\begin{figure*}[!htbp]
 \centering
 \includegraphics[width=0.85\textwidth]{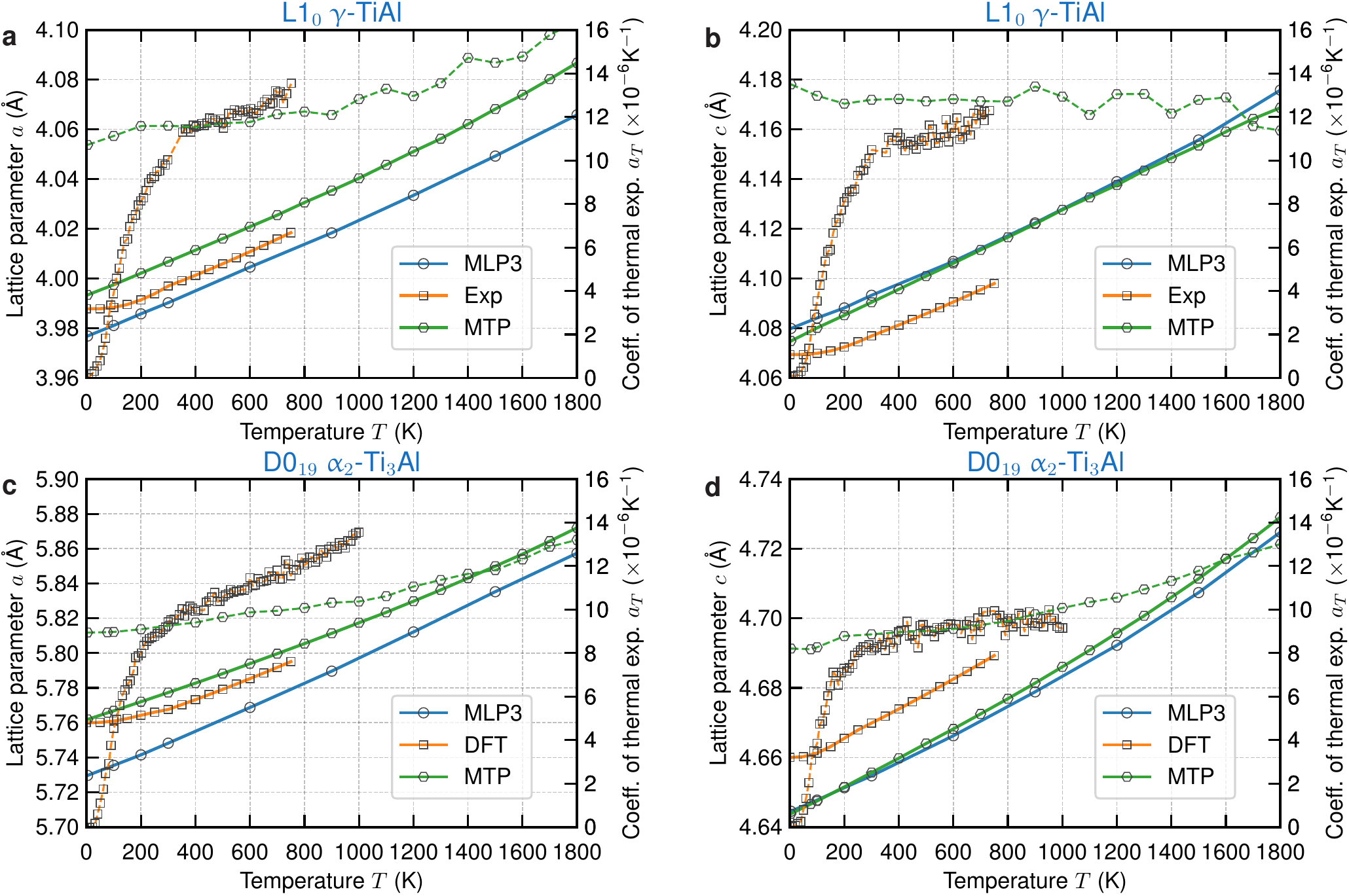}
 \caption{\label{fig:lat_p} Lattice constants and coefficients of thermal expansion of L1\(_0\) \(\gamma\)-TiAl and D0\(_{19}\) \(\alpha_2\)-Ti\(_{3}\)Al as a function of temperature predicted by the MTP and MLP3~\cite{seko_2020_prb} in comparison with experimental~\cite{he_1997_msea} and DFT values~\cite{holec_2019_mat}. (a-b) Lattice parameter \(a\) and \(c\) of L1\(_0\)-TiAl. (c-d) Lattice parameter \(a\) and \(c\) of D0\(_{19}\)-\ce{Ti3Al}. The dashed lines are the coefficients of thermal expansion \(\alpha_T\). The experimental values are properties of a L1\(_0\)-Ti\(_{44}\)Al\(_{56}\) alloy~\cite{he_1997_msea}. }
\end{figure*}

Figure~\ref{fig:lat_p} shows the lattice constants (\(a\) and \(c\)) of the L1\(_0\)-TiAl and D0\(_{19}\)-Ti\(_{3}\)Al structures as a function of temperature. The MTP has accurate lattice parameters for both structures; the differences are within 0.5\% with respect to experimental and DFT values. The MLP3 has comparable accuracies as well. Both IAPs also exhibit normal thermal expansion throughout the entire temperature range. Above room temperature, the two IAPs have thermal expansion coefficients that are in agreement with experiment/DFT values.  Nevertheless, both IAPs have nearly constant coefficients of thermal expansions \(\alpha_T\), in contrast to \(\alpha_T \rightarrow 0\) as \(T \rightarrow 0\)~K seen in experiments and DFT calculations. The diminishing of \(\alpha_T\) at low temperatures arises from quantum effects, while MD with IAPs follow classical mechanics where the vibrational spectra should only be valid above the Debye temperature~\cite{buda_1990_prb,quong_1997_prb}.  This unphysical behaviour is thus common in nearly all empirical IAPs~\cite{qin_2022_cms,wang_2022_prm},  independent of the training datasets.

\begin{figure*}[!htbp]
 \centering
 \includegraphics[width=0.80\textwidth]{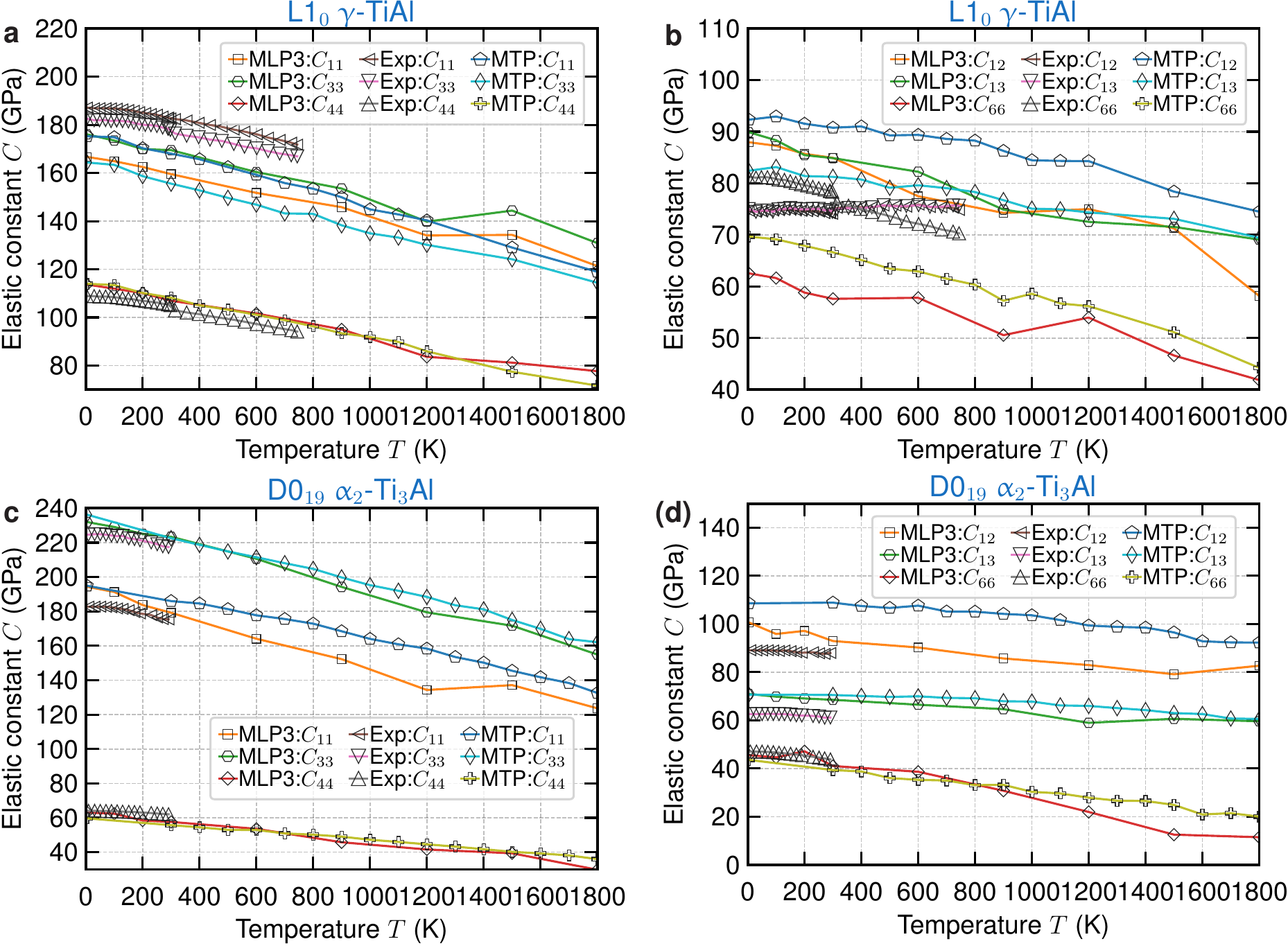}
 \caption{\label{fig:elastic_c_vs_T} Elastic constants of the L1\(_0\) \(\gamma\)-TiAl and D0\(_{19}\) \(\alpha_2\)-Ti\(_{3}\)Al as a function of temperature predicted by the MTP and MLP3~\cite{seko_2020_prb} in comparison with experimental values~\cite{tanaka_1996_pml,he_1997_msea,tanaka_1996_pma}. The discrepancies at finite temperatures are largely inherited from their 0K offsets. }
\end{figure*}

Figure~\ref{fig:elastic_c_vs_T} shows the elastic constants of the L1\(_0\)-TiAl and D0\(_{19}\)-\ce{Ti3Al} structures as a function of temperature.  For both ML-IAPs, all elastic constants decrease gradually with increasing temperatures.  The largest discrepancies lie in the values of \(C_{12}\) of the L1\(_0\)-TiAl and D0\(_{19}\)-\ce{Ti3Al} structures.  The large discrepancies are also present at 0~K and are thus likely inherited from the datasets generated in DFT calculations; it is not uncommon to have discrepancies exceeding 15\% in elastic constants prediction by DFT with respect to experiments (see Table~\ref{tab:lat_prop_0K}).  Nonetheless, the MTP has much improved overall accuracy in reproducing the elastic constants of both intermetallic structures relative to all other IAPs.

\subsection{Surface energies}
\label{sec:surf_e_result}

\begin{figure*}[!htbp]
 \centering
  \includegraphics[width=0.8\textwidth]{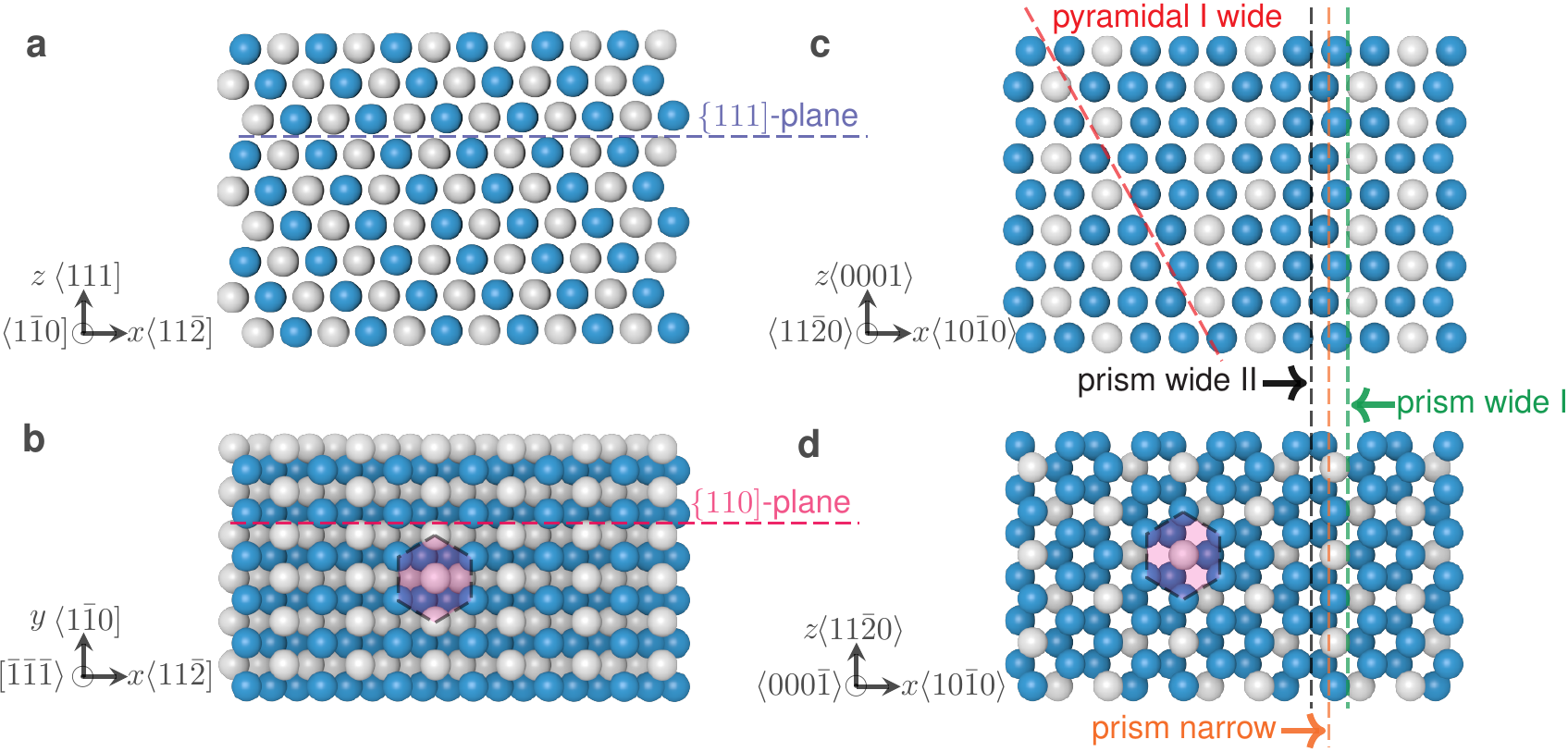}
  \caption{\label{fig:xtal_plane} Atomic planes in the L1\(_0\) \(\gamma\)-TiAl and D0\(_{19}\) \(\alpha_2\)-\ce{Ti3Al} structures. (a-b) The \(\{111\}\) surface terminated by Ti (blue) and Al (white) atoms. (c-d) The prism plane has 3 types of atom-planes: (i) prism wide I terminated by Ti and Al atoms; (ii) prism wide II terminated by Ti atoms; (iii) prism narrow terminated by Ti and Al atoms. The pyramidal I wide plane is terminated by Ti and Al atoms. }
\end{figure*}

Table~\ref{tab:surf_e} shows the surface energies \(\gamma_\text{surf}\) calculated by the IAPs in comparison with that by DFT. We focus on the low index, non-polar planes (Fig.~\ref{fig:xtal_plane}).  In the L1\(_0\) structure, the \(\{111\}\) and \(\{100\}\) plane is terminated by a combination of Ti and Al atoms at equal atomic composition.  The MTP has its \(\gamma_\text{surf}^{\{111\}}\) within 2\% from the DFT value, while the MLP3 substantially underestimates \(\gamma_\text{surf}^{\{111\}}\) by 48\%.  Both classical IAPs also underestimate this surface energy by 28\% and 11\%, respectively.  For the \(\{100\}\) plane, the MTP overestimates \(\gamma_\text{surf}^{\{100\}}\) by 7\%, in comparison with the \(-67\)\%, \(-27\)\% and \(17\)\% of the MLP3, EAM and MEAM potentials, respectively.

\begin{table*}[!htbp]
  \caption{ Surface energies \(\sigma_\text{surf}\) (J/m\(^2\)) predicted by 4 interatomic potentials and DFT. The percentage errors with respect to DFT values are shown in the parenthesis and errors exceeding \(\pm\)15\% are highlighted in bold font. The mean percentage error is defined as \(\overline{\Delta \gamma}^\text{IAP}_\text{surf} = 1/n\sum_i^n{ \lvert \gamma_\text{i}^\text{IAP} - \gamma_\text{i}^\text{DFT} \rvert}/\gamma_\text{i}^\text{DFT} \), where the summation is carried out over all the low index planes examined. See Fig.~\ref{fig:xtal_plane} for the crystallographic planes of the respective surfaces. }
 \small
 \centering
  \begin{tabular}{ll l lllll}
\toprule
\textbf{Structure} & \textbf{Surface} & \textbf{DFT} & \textbf{MTP} & \textbf{MLP3}~\cite{seko_2020_prb} & \textbf{EAM}~\cite{zope_2003_prb} & \textbf{MEAM}~\cite{kim_2016_cms} \\
   \hline
\multirow{2}{*}{L1\(_0\) TiAl}
& \(\{111\}\) & 1.667 & 1.632 (\(-\)2\%) & 0.874 (\(-\)\textbf{48\%}) & 1.193 (\textbf{\(-\)28\%}) & 1.481 (\(-\)11\%) \\

          &  \(\{100\}\)  & 1.643 & 1.755 (7\%) & 0.537 (\(-\)\textbf{67\%}) & 1.206 (\textbf{\(-\)27\%}) & 1.929 (\textbf{17\%}) \\
          & \(\overline{\Delta \gamma}^\text{IAP}_\text{surf}\) & & 5\% & \(-\)\textbf{58\%} & \(-\)\textbf{28\%} & 14\% \\
   \\
\multirow{7}{*}{D0\(_{19}\) \ce{Ti3Al}}
& \(\{0001\}\) & 1.958 & 1.804 (\(-\)8\%) & 0.615 (\(-\)\textbf{69}\%) & 1.258 (\textbf{\(-\)36\%}) & 1.887 (\(-\)4\%) \\
& \(\{1\bar{1}00\}_\text{wide I}\) & 1.940 & 1.890 (\(-\)3\%) & 0.810 (\textbf{\(-\)58\%}) & 1.419 (\textbf{\(-\)27\%}) & 1.797 (\(-\)7\%) \\
& \(\{1\bar{1}00\}_\text{wide II}\) & 1.989 & 1.924 (\(-\)3\%) & 0.534 (\textbf{\(-\)73\%}) & 1.428 (\textbf{\(-\)28\%}) & 2.030 (2\%) \\
& \(\{1\bar{1}00\}_\text{narrow}\) & 2.354 & 2.323 (\(-\)1\%) & 1.191 (\textbf{\(-\)49\%}) & 1.828 (\textbf{\(-\)22\%}) & 2.534 (8\%) \\
& \(\{2\bar{1}01\}_\text{wide}\) & 1.912 & 1.925 (1\%) & 0.621 (\textbf{\(-\)68\%}) & 1.444 (\textbf{\(-\)24\%}) & 2.160 (13\%) \\
& \(\{11\bar{2}1\}\) & 2.128 & 2.152 (1\%) & 1.026 (\textbf{\(-\)52\%}) & 1.638 (\textbf{\(-\)23\%}) & 2.218 (4\%) \\
& \(\overline{\Delta \gamma}^\text{IAP}_\text{surf}\) & & 3\% & \(-\)\textbf{62\%} & \(-\)\textbf{27\%} & 6\% \\
\\
\multirow{5}{*}{Ti}
& \(\{0001\}_\text{basal}\) & 1.968 & 1.716 (\(-\)13\%) & \(-\)0.034 (\textbf{\(-\)102\%}) & 1.275 (\textbf{\(-\)35\%}) & 2.151 (9\%) \\
& \(\{10\bar{1}0\}_\text{pris I}\) & 1.993 & 1.971(\(-\)1\%) & 0.082 (\textbf{\(-\)96\%}) & 1.507 (\textbf{\(-\)24\%}) & 2.369 (\textbf{19\%}) \\
& \(\{1100\}_\text{pris II}\) & 2.483 & 2.328 (\(-\)6\%) & 0.540 (\textbf{\(-\)78\%}) & 1.865 (\textbf{\(-\)25\%}) & 2.471 (0\%) \\
& \(\{10\bar{1}1\}_\text{pyra I}\) & 1.973 & 1.979 (0\%) & 0.170 (\textbf{\(-\)91\%}) & 1.544 (\textbf{\(-\)22\%}) & 2.438 (\textbf{24\%}) \\
& \(\{11\bar{2}2\}_\text{pyra II}\) & 1.967 & 2.000 (2\%) & 0.230 (\textbf{\(-\)88\%}) & 1.579 (\textbf{\(-\)20\%}) & 2.423 (\textbf{23\%}) \\
          & \(\overline{\Delta \gamma}^\text{IAP}_\text{surf}\) & & 4\% & \(-\)\textbf{91\%} & \(-\)\textbf{25\%} & \textbf{15\%} \\
   \\
\multirow{3}{*}{Al}
& \(\{111\}\) & 0.834 & 0.717 (\(-\)14\%) & 0.601 (\textbf{\(-\)28\%}) & 0.596 (\textbf{\(-\)29\%}) & 0.626 (\textbf{\(-\)25\%}) \\
& \(\{110\}\) & 0.982 & 0.907 (\(-\)8\%) & 0.898 (\(-\)8\%) & 0.787 (\textbf{\(-\)20\%}) & 0.919 (\(-\)6\%) \\
& \(\{100\}\) & 0.933 & 0.838 (\(-\)10\%) & 0.756 (\textbf{\(-\)19\%}) & 0.605 (\textbf{\(-\)35\%}) & 0.847 (\(-\)9\%) \\
          & \(\overline{\Delta \gamma}^\text{IAP}_\text{surf}\) & & \(-\)11\% & \(-\)\textbf{18\%} & \(-\)\textbf{28\%} & 13\% \\
\bottomrule
  \end{tabular}
 \label{tab:surf_e}
\end{table*}

In the D0\(_{19}\) \ce{Ti3Al} structure, the IAPs have similar accuracy trends as that in the L1\(_0\) structure.  The MTP has accurate \(\gamma_\text{surf}\) in all the surfaces examined; its largest discrepancy is \(-\)8\% in \(\gamma_\text{surf}^{\{0001\}}\) on the basal plane. The MLP3 severely underestimates \(\gamma_\text{surf}\) by roughly 2/3 with respect to DFT values on nearly all the surfaces.  The classical EAM potential~\cite{zope_2003_prb} also underestimates \(\gamma_\text{surf}\) by \(\sim\)1/4, while the MEAM potential~\cite{kim_2016_cms} accurately reproduces \(\gamma_\text{surf}\) with its largest discrepancy at 13\%. Furthermore, the general trends in \(\gamma_\text{surf}\) are also seen in the elemental structures of HCP Ti and FCC Al; the MTP accurately reproduces all the \(\gamma_\text{surf}\) and the MLP3 underestimates all the \(\gamma_\text{surf}\).  In addition, the overall relative accuracy, as measured by the mean percentage error for \(n\) planes defined as \(\overline{\Delta \gamma}^\text{IAP}_\text{surf} = 1/n\sum_i^n{ \lvert \gamma_\text{i}^\text{IAP} - \gamma_\text{i}^\text{DFT} \rvert}/\gamma_\text{i}^\text{DFT} \), follows the same trends among the different phases in all the IAPs.  The overall data on \(\gamma_\text{surf}\) suggests that explicit inclusion of surface structures in the training dataset has substantially improved the reproducibility of surface energy in the MTP.

\begin{figure*}[!htbp]
 \includegraphics[width=0.8\textwidth]{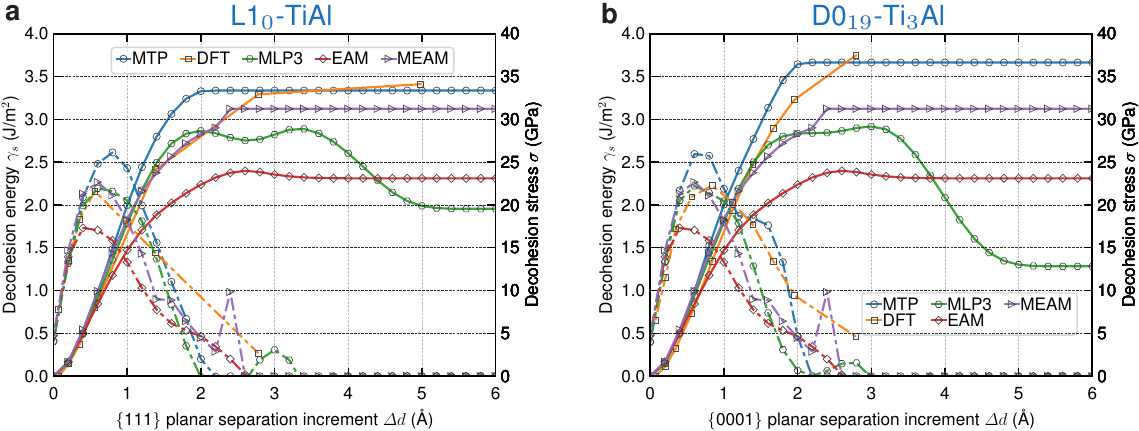}
 \caption{\label{fig:surface_decoh} Separation-cohesion relation during surface decohesion along two atomic planes predicted by 4 interatomic potentials and DFT. (a) \(\{111\}\) plane in the L1\(_0\) \(\gamma\)-TiAl. (b) \(\{0001\}\) plane in the D0\(_{19}\) \(\alpha_2\)-\ce{Ti3Al}.  The solid and dashed lines are the decohesion energies and stresses, respectively.  The MTP matches well with DFT calculations~\cite{kanani_2014_im} and is much more accurate than the MLP3~\cite{seko_2020_prb}, EAM~\cite{zope_2003_prb} and MEAM~\cite{kim_2016_cms} IAPs. }
\end{figure*}

Figure~\ref{fig:surface_decoh} shows the surface decohesion-separation energy and stress curves of the \(\{111\}\)-plane in TiAl and \(\{0001\}\)-plane in \ce{Ti3Al}.  The MTP accurately reproduces the smooth energy variations during surface decohesion.  It also has an accurate decohesion stress of 25 GPa, which is close to the DFT value of 22 GPa in both cases.  At large planar separations \(\Delta d \in [1, 5]\) \r{A}, the MLP3 has unphysical behaviour where the decohesion energies do not monotonically increase and approach \(2\gamma_\text{surf}\).  Such behaviour is not uncommon in ML-IAPs when the surface decohesion information is not included in the potential training dataset; this again demonstrates the transferability of ML-IAPs are not always guaranteed automatically.  The surface decohesion behaviour of the classical IAPs have been studied earlier; the EAM potential underestimates \(\gamma_\text{surf}\) and the decohesion stress, while the MEAM exhibits discontinuity in the energy variation and spurious decohesion stress (see Ref.~\cite{pei_2021_cms}).  The MTP is thus the only IAP capable of modelling cleavage processes in the Ti-Al system.

\subsection{Generalized stacking fault energy}

\begin{figure*}[!htbp]
 \centering
 \includegraphics[width=0.90\textwidth]{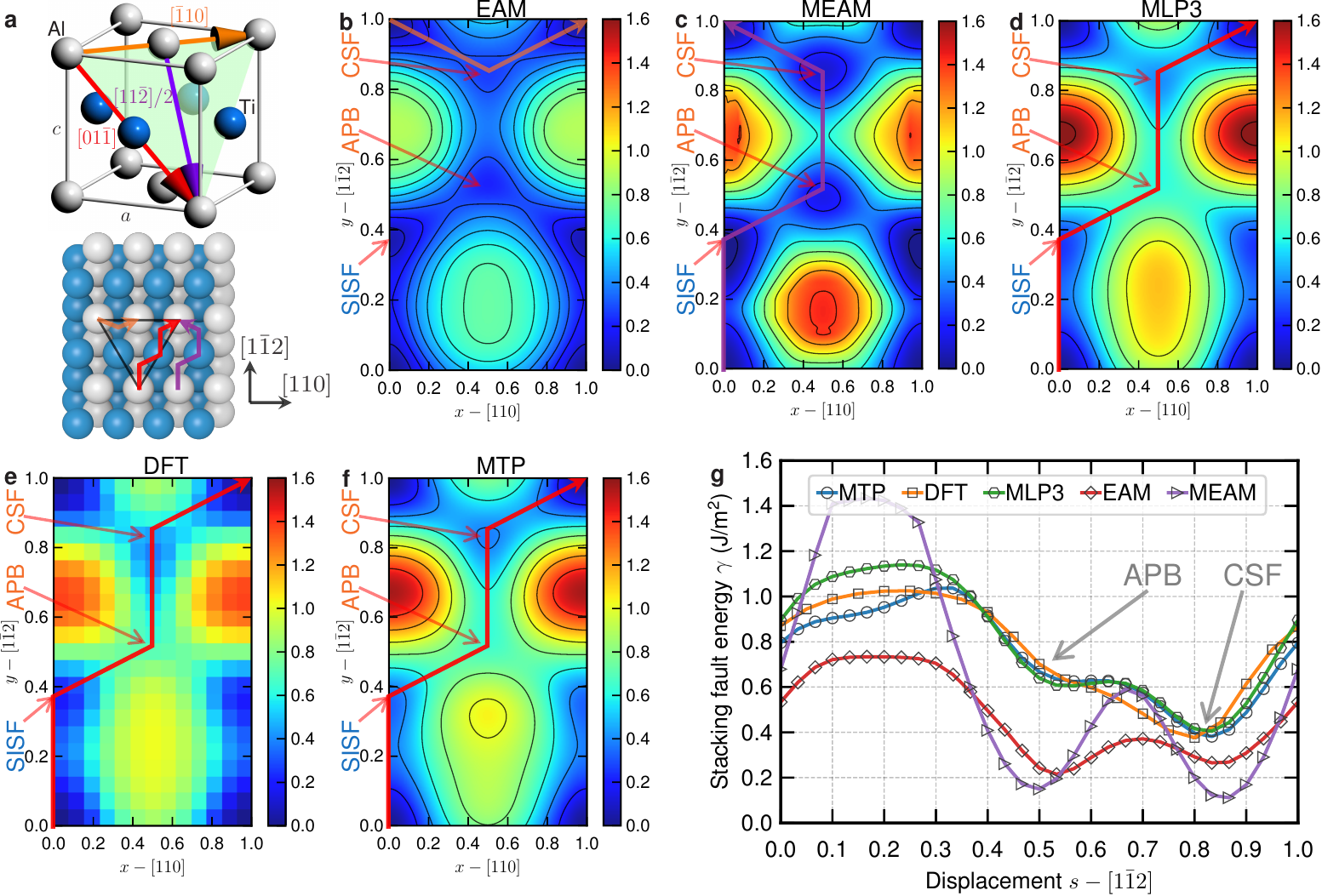}
 \caption{\label{fig:gamma_surf_gamma} Generalized stacking fault energy of the \(\{111\}\) plane in the L1\(_0\) \(\upgamma\)-TiAl structure predicted by 4 interatomic potentials and DFT. (a) The L1\(_0\) unit cell and \(\{111\}\)-plane slip system. (b) EAM~\cite{zope_2003_prb}. (c) MEAM~\cite{kim_2016_cms}. (d) MLP3~\cite{seko_2020_prb}. (e) DFT. (f) MTP. (g) The \(\gamma\)-line along the \([1\bar{1}2]\) direction at slip \(\mathbf{s}=[110]/2\) passing through the APB and CSF. The orange and red lines are the slip paths for the ordinary \(\langle \bar{1}10 ]/2 \) and \(\langle 01\bar{1}]\) super dislocations. All \(\gamma\)-surfaces here and in subsequent figures are calculated with structures optimized in the slip-plane normal (\(z\)) direction and atoms fixed in the in-plane (\(x-y\)) directions. }
\end{figure*}

We further examine the GSFE \(\gamma\)-surfaces in the two structures. Figure~\ref{fig:gamma_surf_gamma} shows the \(\gamma\)-surface of the \(\{111\}\) plane in the L1\(_0\) \(\upgamma\)-TiAl structure. The MTP reproduces the entire \(\gamma\)-surface in agreement with DFT calculations, so does the MLP3. However, the classical IAPs show both quantitative and qualitative discrepancies from the \(\gamma\)-surface calculated by DFT. In particular, the EAM potential~\cite{zope_2003_prb} substantially underestimates the overall \(\gamma\)-surface while the MEAM potential~\cite{kim_2016_cms} has a different energy profile from the rest.  We note that the MEAM potential and the MTP have relatively short cutoff radii of $r_\text{c}$ = 5.0 \AA\ and $4.8$ \AA\, while the EAM and MLP3 have $r_\text{c}$ = 6.7 \AA\ and 8.0 \AA; the latter two IAPs have the best qualitative match in terms of the general profile of the \(\{111\}\)-plane \(\gamma\)-surface, suggesting the importance of appropriate interatomic cutoff radius in reproducing defect properties.

On the \(\{111\}\)-plane \(\gamma\)-surface, three SFs are relevant for dislocation core dissociations and structures. The CSF governs the ordinary \(\langle 110 ]/2\) dislocation while the SISF and APB control the \(\langle 10\bar{1} ]\) and \(\langle 11\bar{2}]/2\) superdislocations (detailed below). Figure~\ref{fig:gamma_surf_gamma}g shows the \(\gamma\)-lines of all the models along the \(\langle 1\bar{1}2]\) direction, which passes through the APB and CSF. The MTP and MLP3 reproduce the APB and CSF energies within 10\% from the DFT values, while the two classical IAPs substantially underestimate both the APB and CSF energies (Table~\ref{tab:sf_e}). DFT predicts that the APB is unstable in the \(\langle 1\bar{1}2\rangle\) direction; the two ML-IAPs exhibit a similar feature with a very shallow local energy minimum at the APB position, in contrast to the deep meta-stable APB energies of the two classical IAPs. In addition, the MEAM potential~\cite{kim_2016_cms} has a negative SISF energy of \(-\)77 mJ/m\(^2\), while both the MTP and MLP3 overestimate the SISF energy. The negative SISF energy in the MEAM potential suggests that the L1\(_0\) phase may be unstable against some other phases not examined in Fig.~\ref{fig:form_energy}. Table~\ref{tab:sf_e} also shows various SF energies calculated previously by first-principles methods; considerable discrepancies exist among these data. However, all calculations show the same trend of SF energy ordering in the L1\(_0\) TiAl structure, i.e., \(\gamma_\text{SISF} < \gamma_\text{CSF} < \gamma_\text{APB}\). This ordering is also reproduced by the two ML-IAPs.

\begin{table*}[!htbp]
  \caption{Stacking fault energies \(\gamma_\text{sf}\) (mJ/m\(^2\)) predicted by 4 interatomic potentials and DFT.  In L1\(_0\)-TiAl, the APB is unstable in DFT and the reported value is taken at slip \(\mathbf{s}_x=0.5[110], \mathbf{s}_y=0.55[1\bar{1}2]\) (Fig.~\ref{fig:gamma_surf_gamma}).  The percentage errors are calculated with respect to the DFT values obtained in this work and are shown in the parenthesis. Errors exceeding \(\pm\)15\% are highlighted in bold font. }
 \small
 \centering
 \begin{tabular}{l rr p{5cm} rrrrr}
  \toprule
\textbf{Structure} & \textbf{Plane} & \textbf{Stacking fault} & \textbf{DFT} & \textbf{MTP} & \textbf{MLP3}~\cite{seko_2020_prb} & \textbf{EAM}~\cite{zope_2003_prb} & \textbf{MEAM}~\cite{kim_2016_cms} \\
  \hline
  \multirow{3}{*}{L1\(_0\)-TiAl}& \multirow{3}{*}{\(\{111\}\)} & SISF & 182, 123~\cite{woodward_1996_pma}, 90~\cite{yoo_1998_mmta}, 160~\cite{woodward_2004_pm}, 133~\cite{liu_2007_im}, 178~\cite{jeong_2018_sr}, 177~\cite{kanani_2014_im}, 194~\cite{seko_2020_prb} & 322 (\textbf{77\%}) & 242 (\textbf{33\%}) & 57 (\(-\)\textbf{69\%}) & \(-\)77 (\(-\)\textbf{142\%}) \\
& & APB & 600, 672~\cite{woodward_1996_pma}, 560~\cite{yoo_1998_mmta}, 610~\cite{woodward_2004_pm}, 641~\cite{jeong_2018_sr}, 603~\cite{kanani_2014_im}, 681~\cite{seko_2020_prb} & 611 (1.8\%) & 592 (\(-\)1.3\%) & 213 (\(-\)\textbf{65\%}) & 146 (\(-\)\textbf{76\%}) \\
          & & CSF & 356, 294~\cite{woodward_1996_pma}, 410~\cite{yoo_1998_mmta}, 372~\cite{woodward_2004_pm}, 352~\cite{jeong_2018_sr}, 371~\cite{kanani_2014_im}, 388~\cite{seko_2020_prb} & 372 (4.5\%) & 393 (10\%) & 261 (\(-\)\textbf{27\%}) & 94 (\(-\)\textbf{74\%}) \\
 \\
\multirow{7}{*}{D0\(_{19}\)-\ce{Ti3Al}} & \multirow{3}{*}{\(\{0001\}\)} & SISF & 93, 117~\cite{liu_2007_im} & 84 (\(-10\)\%) & 100 (\(8\)\%) & 4 (\(-\)\textbf{96\%}) & 225 (\textbf{142\%}) \\
& & APB & 256, 257~\cite{koizumi_2006_pm} & 213 (\(-\)\textbf{17\%}) & 320 (\textbf{25\%}) & 93 (\(-\)\textbf{64\%}) & 163 (\(-\)\textbf{36\%}) \\
& & CSF & 320 & 309 (\(-3\)\%) & 255 (\(-\)\textbf{20\%}) & 96 (\(-\)\textbf{70\%}) & 323 (1\%) \\
  \\
          & \multirow{3}{*}{\(\{1\bar{1}00\}\)} & APB narrow & 82 & 84 (2.4\%) & 238 (\textbf{190\%}) & 49 (\(-\)\textbf{40\%}) & 182 (\textbf{122\%}) \\
& &  APB Wide I & 464 & 487 (5\%) & 538 (\textbf{16\%}) & 181 (\(-\)\textbf{61\%}) & 173 (\(-\)\textbf{63\%}) \\
          & & APB wide II & 82 & 84 (2.4\%) & 238 (\textbf{190\%}) & 49 (\(-\)\textbf{40\%}) & 182 (\textbf{122\%}) \\
  \\
& \(\{2\bar{2}01\}\) & APB Wide & 230 & 220 (\(-\)4\%) & 399 (\textbf{73\%}) & 128 (\(-\)\textbf{44\%}) & 184 (\(-\)\textbf{20\%}) \\
  \\
& \(\{11\bar{2}1\}\) & APB & 263 & 243 (\(-\)8\%) & 395 (\textbf{50\%}) & \(-\)125 (\(-\)\textbf{148\%}) & \(-\)12 (\(-\)\textbf{105\%})
  \\
  \bottomrule
 \end{tabular}
 \label{tab:sf_e}
\end{table*}

The D0\(_{19}\) \ce{Ti3Al} phase is complex as it has multiple slip systems on different crystallographic planes (Fig.~\ref{fig:unit_cell_gamma_alpha2}). The \(\gamma\)-surfaces of the D0\(_{19}\) structure have not been calculated at DFT accuracies previously; such calculations require large supercells and thus considerable computing resources. In the present work, we have computed the \(\gamma\)-surfaces on the basal and prism slip planes and all the relevant \(\gamma\)-lines on the basal, prism, pyramidal I and II planes using DFT. With these DFT results, quantitative comparisons are made among the various IAPs.

\begin{figure*}[!htbp]
 \centering
 \includegraphics[width=0.995\textwidth]{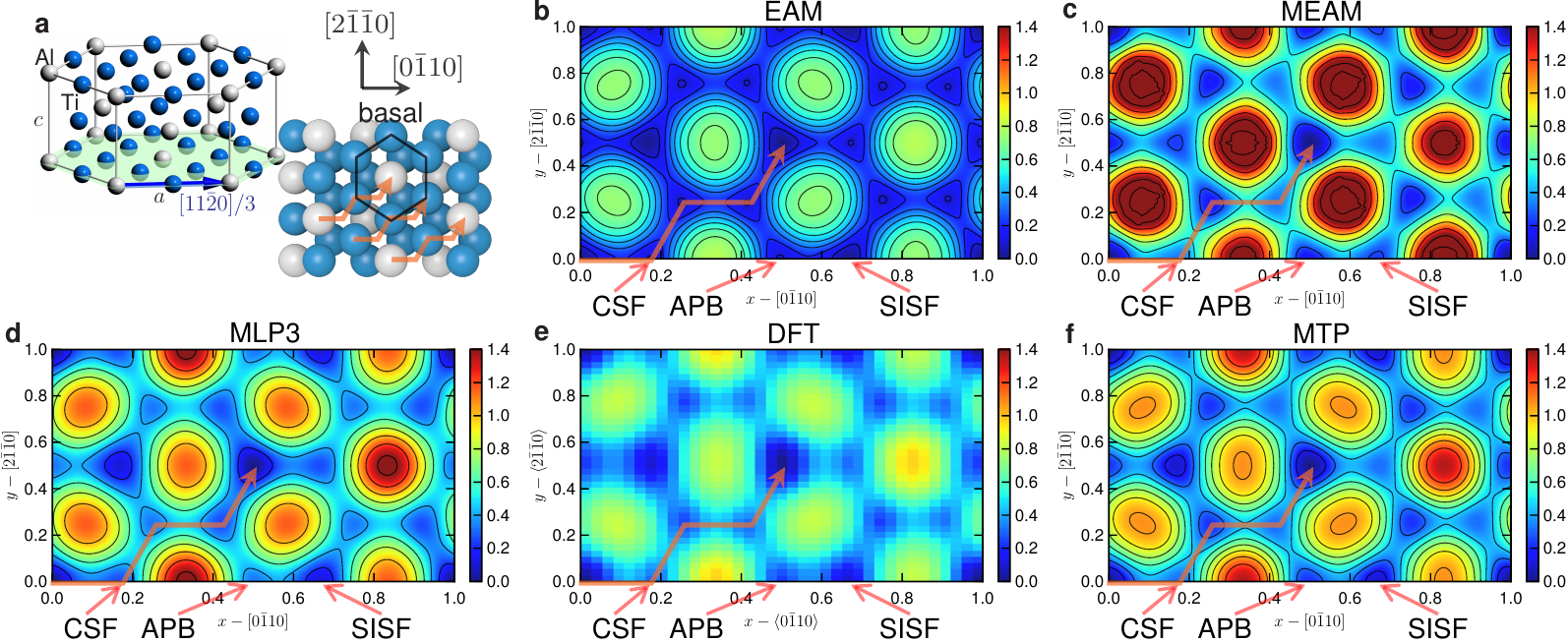}
 \caption{\label{fig:gamma_surf_basal_alpha2} Generalized stacking fault energy surface of the \(\{0001\}\) plane in the D0\(_{19}\) \(\upalpha_2\)-\ce{Ti3Al} structure. (a) The D0\(_{19}\) unit cell and slip path on the basal plane. (b) EAM~\cite{zope_2003_prb}. (c) MEAM~\cite{kim_2016_cms}. (d) MLP3~\cite{seko_2020_prb}. (e) DFT. (f) MTP. In (b-f), the red arrow is the slip path passing through the CSF and APB. The zig-zag path is the expected slip path and highlighted by the red arrows.}
\end{figure*}

\begin{figure}[!htbp]
 \centering
 \includegraphics[width=0.5\textwidth]{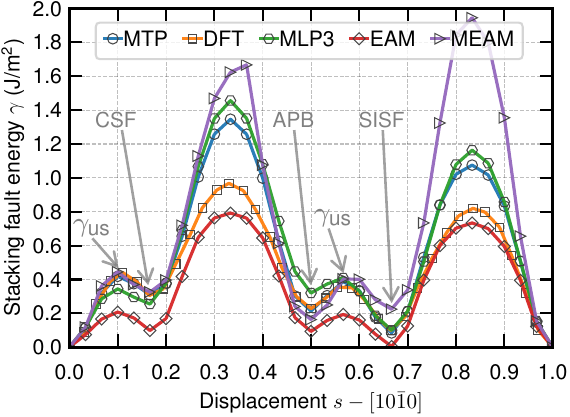}
 \caption{\label{fig:gamma_line_basal_alpha2} Generalized stacking fault energy line along the \(\langle 10\bar{1}0 \rangle\) direction on the \(\{0001\}\) plane in the D0\(_{19}\) \(\upalpha_2\)-\ce{Ti3Al} structure (see Fig.~\ref{fig:gamma_surf_basal_alpha2}). }
\end{figure}

Figures~\ref{fig:gamma_surf_basal_alpha2} and~\ref{fig:gamma_line_basal_alpha2} show the \(\gamma\)-surface and \(\gamma\)-line in the \([10\bar{1}0]\) direction on the basal plane. The MTP reproduces the general \(\gamma\)-surface profile in agreement with DFT. In particular, it has the SISF, APB and CSF energies close to the DFT values (Table~\ref{tab:sf_e}). The MLP3 also has relatively accurate SF energies on the basal plane; it overestimates the APB energy by 25\% and underestimates the CSF energy by 20\%. In addition, the MTP also reproduces the unstable SF energies \(\gamma_\text{us}\) at slips \(\mathbf{s}=0.1 \langle 1\bar{1}00 \rangle\) and \(\mathbf{s}=0.56 \langle 1\bar{1}00 \rangle\). These unstable SFs govern the dislocation nucleation barriers on the basal plane. On the other hand, the EAM potential substantially underestimates all SF energies and has nearly zero SISF energy at slip \(\mathbf{s}=0.67\langle 1\bar{1}00 \rangle\), which will lead to an unrealistically large partial dislocation separation on the \(\{0001\}\) plane. The MEAM potential overestimates the SISF energy by 142\% and underestimates the APB energy by 36\%.  While the MTP and MLP3 overestimate the unstable SF energy at slips \(\mathbf{s}=0.33 \langle 1\bar{1}00 \rangle\) and \(\mathbf{s}=0.83 \langle 1\bar{1}00 \rangle\), these SFs do not lie along the slip path on the basal plane and thus do not control plastic slip (Fig.~\ref{fig:gamma_surf_basal_alpha2}). Among all the IAPs, the MTP has the best agreement with DFT on the basal plane.

\begin{figure*}[!htbp]
  \centering
  \includegraphics[width=\textwidth]{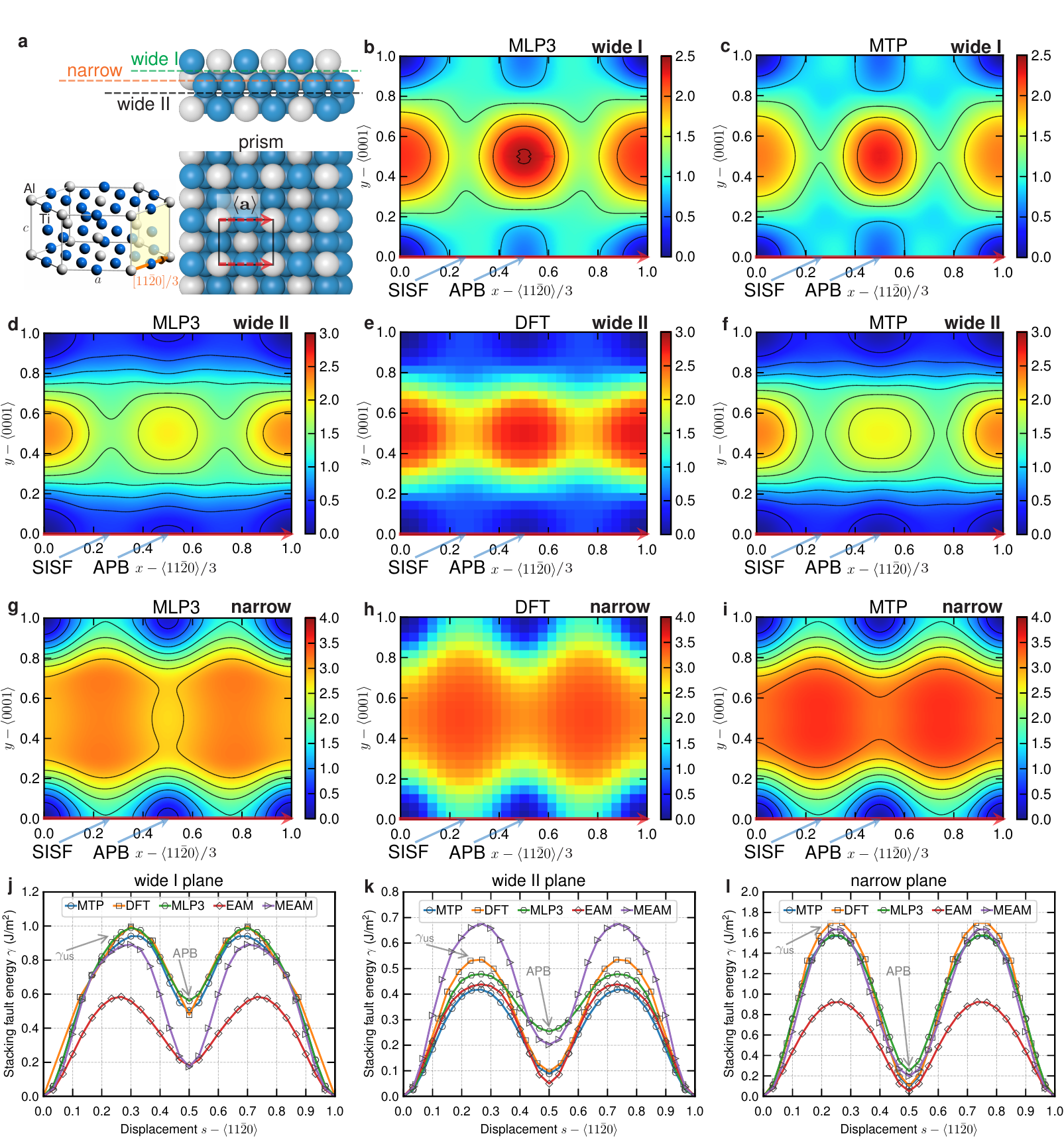}
  \caption{\label{fig:gamma_surf_prism_alpha2} Generalized stacking fault energy of the \(\{10\bar{1}0\}\) prism plane in the D0\(_{19}\) \(\alpha_2\)-\ce{Ti3Al} structure predicted by 4 interatomic potentials and DFT. (a) The prism plane in the D0\(_{19}\) unit cell. (b-c) \(\gamma\)-surfaces of the prism wide I plane by the MLP~\cite{seko_2020_prb} and MTP potential. (d-f) \(\gamma\)-surfaces of the prism wide II plane by the MLP~\cite{seko_2020_prb}, DFT and MTP. (g-i) \(\gamma\)-surfaces of the prism narrow plane by the MLP~\cite{seko_2020_prb}, DFT and MTP. The red arrow is the slip path passing through the APB. (j-l) The \(\gamma\)-lines along the \(\langle 11\bar{2}0 \rangle\) slip direction on the 3 \(\{10\bar{1}0\}\) plane in the \(\upalpha_2\)-\ce{Ti3Al} structure calculated by 4 interatomic potentials and DFT. }
\end{figure*}

On the \(\{1\bar{1}00\}\) prism plane, three slip planes exist depending on their interplanar distance and site occupancy (Fig.~\ref{fig:xtal_plane}c-d). Slip can occur along (i) the narrow plane between two narrowly-spaced atom layers of Ti, and both Ti and Al; (ii) the wide I plane between two widely-spaced atom layers of both Ti and Al and (iii) the wide II plane between two widely-spaced atom layers of Ti only (Fig.~\ref{fig:xtal_plane}). Figure~\ref{fig:gamma_surf_prism_alpha2} shows the \(\gamma\)-surface and \(\gamma\)-line in the \([11\bar{2}0]\) direction on the prism plane. On the narrow plane, the two ML-IAPs and the MEAM IAP are able to reproduce the general \(\gamma\)-line profile. The MTP has an accurate APB energy deviating only 2.4\% from the DFT value, while the MLP3 and the MEAM IAP overestimate the APB energy by 190\% and 122\%, respectively (Table~\ref{tab:sf_e}). The EAM IAP substantially underestimates the entire \(\gamma\)-line; both its unstable SF energy and \(\gamma_\text{APB}\) are only \(\sim\)50\% of the DFT values.

On the prism wide I plane (Fig.~\ref{fig:gamma_surf_prism_alpha2}j), the EAM IAP again substantially underestimates the \(\gamma\)-line. Both the EAM and MEAM IAPs underestimate the APB energies by \(60\)\% relative to the DFT value. The two ML-IAPs reproduce the general \(\gamma\)-line profile. In particular, the MTP has its APB energy differing by 5\% from the DFT value, while the MLP3 overestimates the APB energy by 16\%. On the prism wide II plane (Fig.~\ref{fig:gamma_surf_prism_alpha2}k), the MEAM IAP overestimates the entire \(\gamma\)-line with its APB energy 122\% higher than the DFT value, while the MLP3 overestimates the APB energy by 190\%. The MTP has an APB energy nearly identical to the DFT value and an unstable SF energy at 20\% lower than the DFT value. We note that the APB on the narrow plane and wide II plane have the same atom configuration and thus an identical APB energy in all models. Among all the IAPs, the MTP again has the best overall \(\gamma\)-lines on the prism slip planes in agreement with DFT. Comparing the \(\gamma\)-lines on all 3 prism planes (Fig.\ref{fig:gamma_surf_prism_alpha2}j-l), the wide II plane, separating 2 Ti-atom planes, has the lowest unstable and stable stacking faults and is expected to be the easy-slip plane on the prism \adisl slip system in D0\(_{19}\)-\ce{Ti3Al}.

\begin{figure*}[!htbp]
  \centering
  \includegraphics[width=\textwidth]{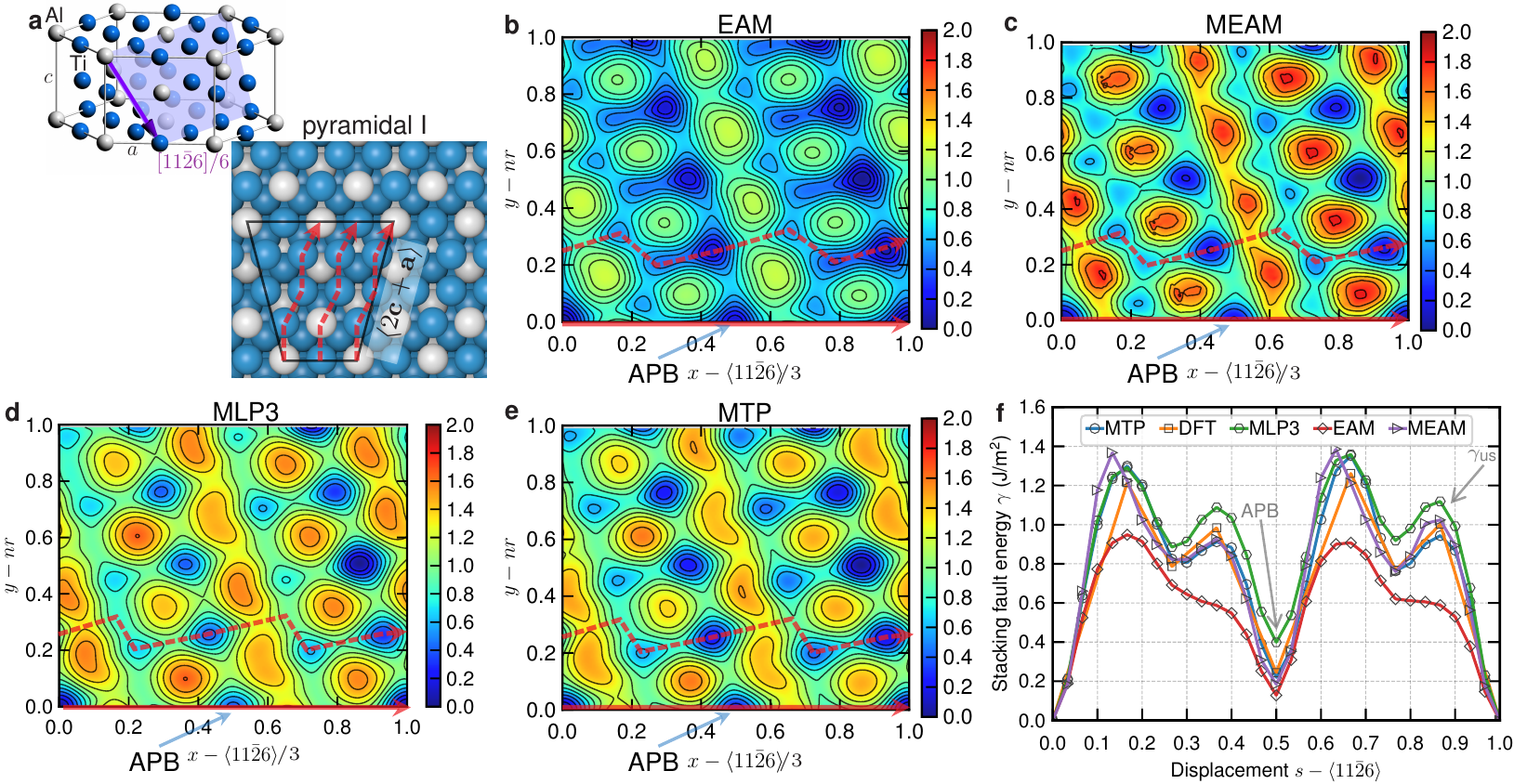}
  \caption{\label{fig:gamma_surf_pyi_alpha2} Generalized stacking fault energy of the \(\{2\bar{2}01\}\) pyramidal I wide plane in the D0\(_{19}\) \(\alpha_2\)-\ce{Ti3Al} structure predicted by 4 interatomic potentials. (a) The pyramidal I slip plane with the \ccadisl slip path highlighted. (b) EAM~\cite{zope_2003_prb}. (c) MEAM~\cite{kim_2016_cms}. (d) MLP3~\cite{seko_2020_prb}. (e) MTP. (f) Comparisons of the \(\gamma\)-lines along the \ccadisl direction. In (b-e), the red solid arrow is the slip path in the \ccadisl passing through the APB. The red dashed arrow is the slip path passing through local meta-stable stacking faults.}
\end{figure*}

Figure~\ref{fig:gamma_surf_pyi_alpha2} shows the \(\gamma\)-surface and \(\gamma\)-line in the \ccadislns-\([11\bar{2}6]\) direction on the \(\{2\bar{2}01\}\) pyramidal I plane. The pyramidal planes are important since plastic strain in the crystallographic \cdisl direction is accommodated by slips in the \(\langle 11\bar{2}6 \rangle\) direction via the \ccadisl dislocations. The EAM and MEAM IAPs have similar \(\gamma\)-surface profiles (Fig.~\ref{fig:gamma_surf_pyi_alpha2}b-c); the EAM IAP exhibits much lower \(\gamma\)-surface and \(\gamma\)-line (Fig.~\ref{fig:gamma_surf_pyi_alpha2}f), as it does on other planes. The MTP and MLP3 IAPs have similar \(\gamma\)-surfaces (Fig.~\ref{fig:gamma_surf_pyi_alpha2}d-e); their \(\gamma\)-surfaces are distinctly different from that of the two classical IAPs. The MTP has an APB energy of 220 mJ/m\(^2\), similar to the DFT value of 230 mJ/m\(^2\), while the MLP3 overestimates the APB energy by 73\%. Furthermore, the MTP captures the various meta-stable and unstable SF energies along the \ccadisl slip direction in agreement with DFT (Fig.~\ref{fig:gamma_surf_pyi_alpha2}f). The MTP is thus the only IAP able to accurately reproduce the entire \(\gamma\)-line on the pyramidal I plane.  In addition, the MEAM, MLP3, MTP and DFT have metastable SF at \(\mathbf{s} \approx 0.25\) \ccadisl and \(\mathbf{s} \approx 0.75\) \ccadisl, which suggest further \ccadisl dislocation dissociations on the pyramidal I plane, as shown in Section~\ref{sec:disl_core} below.

\begin{figure*}[!htbp]
  \centering
  \includegraphics[width=1.0\textwidth]{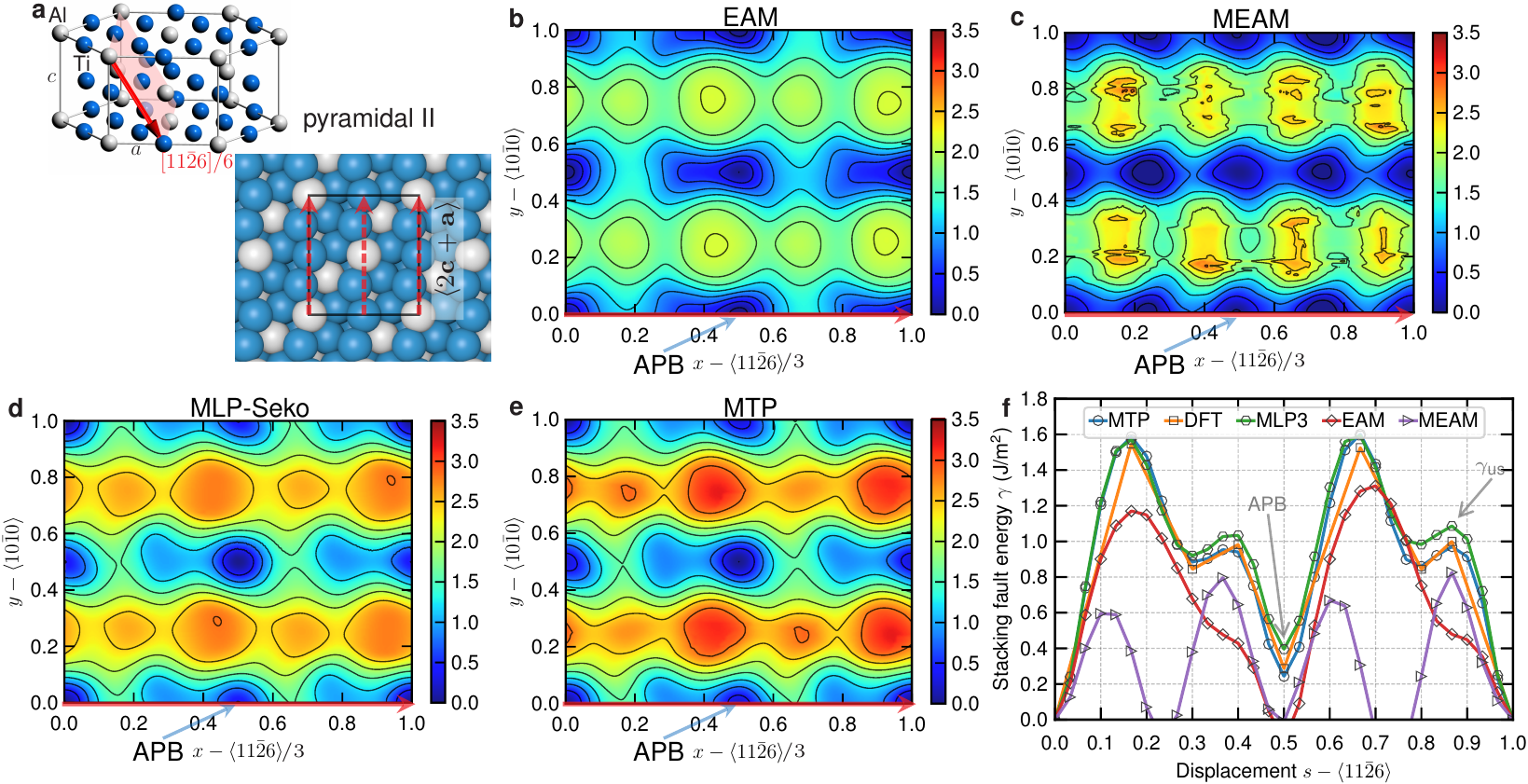}
  \caption{\label{fig:gamma_surf_pyii_alpha2} Generalized stacking fault energy of the \(\{11\bar{2}1\}\) pyramidal II plane in the D0\(_{19}\) \(\alpha_2\)-\ce{Ti3Al} structure. (a) The pyramidal II slip plane with the \ccadisl slip path highlighted. (b) EAM~\cite{zope_2003_prb}. (c) MEAM~\cite{kim_2016_cms}. (d) MLP3~\cite{seko_2020_prb}. (e) MTP. (f) Comparisons of the \(\gamma\)-lines along the \ccadisl direction. In (b-e), the red solid arrow is the slip path in the \ccadisl passing through the APB. }
\end{figure*}

Figure~\ref{fig:gamma_surf_pyii_alpha2} shows the \(\gamma\)-surface and \(\gamma\)-line in the \ccadislns-\([11\bar{2}6]\) direction on the \(\{11\bar{2}1\}\) pyramidal II plane. On this plane, the \(\gamma\)-surface is not entire smooth in the MEAM IAP, which is related to the unsmooth energy at the potential cutoff distance. Both the EAM and MEAM IAPs underestimate the \(\gamma\)-line when compared to DFT (Fig.~\ref{fig:gamma_surf_pyii_alpha2}f). These two classical IAPs also have negative APB energy, in contrast to DFT \(\gamma_\text{APB} = 263 \text{ mJ}/\text{m}^2\) (Table~\ref{tab:sf_e}). The MEAM IAP has negative SF energies at slip \(\mathbf{s} \approx 0.25\)\ccadisl and \(\mathbf{s} \approx 0.75\)\ccadislns, in stark contrast to the high SF energies above \(800 \text{ mJ}/\text{m}^2\) predicted by DFT. The two classical IAPs are thus not appropriate for modelling the \ccadisl dislocations in the D0\(_{19}\) \ce{Ti3Al} structure. On the other hand, both the MLP3 and MTP have similar \(\gamma\)-surface profile and \(\gamma\)-line along the \ccadisl slip direction. Their \(\gamma\)-lines follow closely with that by DFT; they capture nearly all unstable and meta-stable SF energies. At the APB point, the MTP has an APB energy of 243 mJ/m\(^2\), similar to the DFT value of 263 mJ/m\(^2\), while the MLP3 overestimates the APB energy by 50\%. The present MTP thus again clearly outperforms previous IAPs in modelling slip behaviour on the pyramidal II plane of the D9\(_{19}\)-\ce{Ti3Al} intermetallic structure.

Overall, the above comparisons suggest that the two classical IAPs can not reproduce the \(\gamma\)-surfaces and \(\gamma\)-lines in the complex L1\(_0\) \(\gamma\)-TiAl and D0\(_{19}\) \(\alpha_2\)-\ce{Ti3Al} intermetallic structures. The two ML-IAPs share many similar features and have relatively good accuracies when compared to DFT, even though the \(\gamma\)-surface structures are not included in the potential training datasets. In particular, the MTP accurately reproduces nearly all the \(\gamma\)-surfaces and SF energies in agreement with DFT. Given its accuracy on the basic lattice properties and defect energetics, the MTP is appropriate for modelling dislocation and cleavage phenomena in both structures, enabling a wide range of studies to be performed for the first time in the Ti-Al alloy system. Such studies include dislocation dissociation and glide, lattice friction, cross-slip, interface structures and migration, as well as their temperature-dependent behaviour. As a further demonstration of the current MTP, we provide a preliminary study on the dislocation core structures in both phases below.

\subsection{Dislocation Core Structures}
\label{sec:disl_core}

\begin{figure*}[!htbp]
 \centering
 \includegraphics[width=0.55\textwidth]{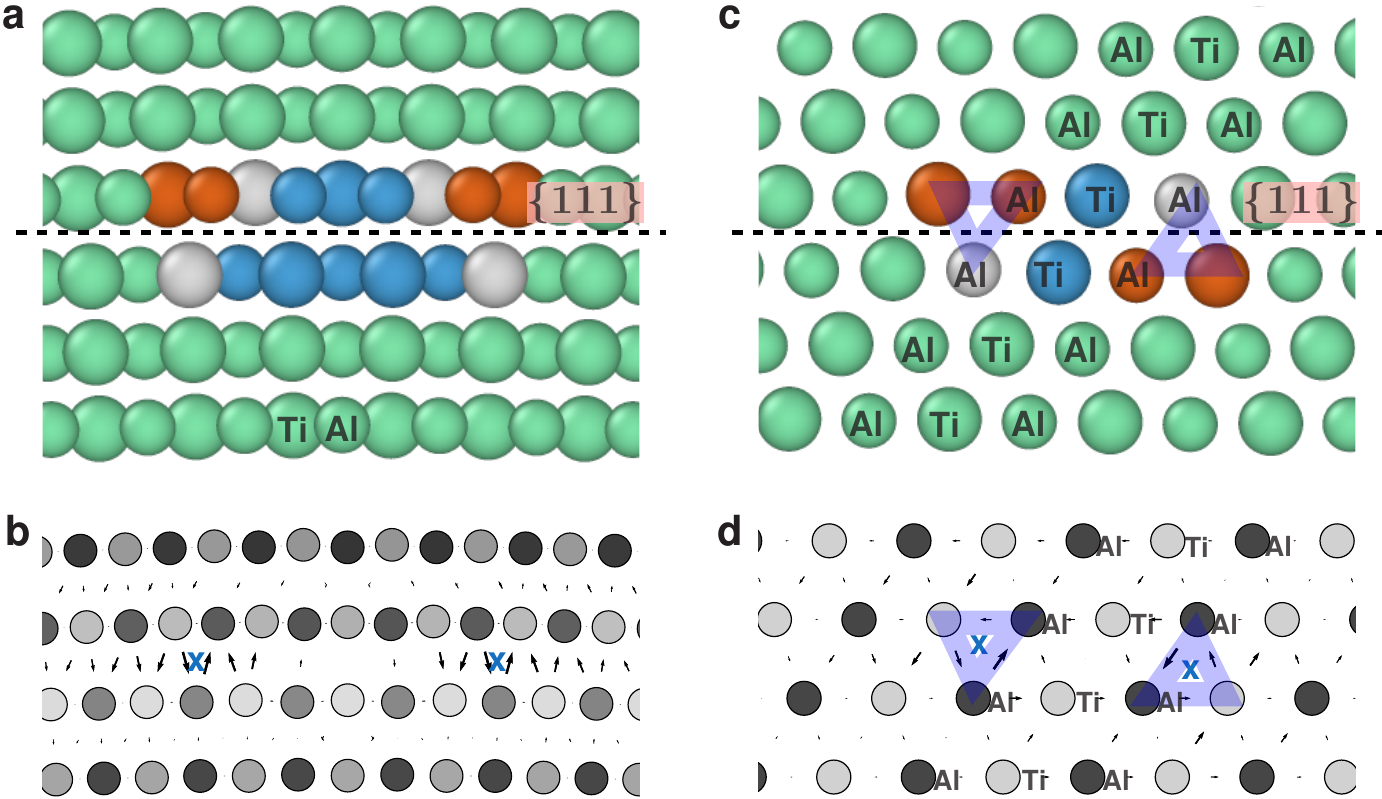}
 \caption{\label{fig:core_a_gamma} The core structures of the \(\langle 110 ]/2\)-\adisl ordinary dislocation on the \{111\} plane in the L1\(_0\) \(\upgamma\)-TiAl structure. (a-b) Edge dislocation. (c-d) Screw dislocation. In (a,c), atoms are colored by local atomic environment identified by the common neighbor analysis (CNA~\cite{faken_1994_cms}): green-FCC, blue-HCP, red-BCC, white-others. In (b,d) the cores are visualized by the differential displacement (DD) map~\cite{vitek_1970_pma}. The same coloring scheme and DD map are used in subsequent figures. }
\end{figure*}

Figure~\ref{fig:core_a_gamma} shows the core structures of the ordinary \adisl dislocation on the \(\{111\}\) slip plane in the \(\upgamma\)-TiAl structure (Fig.~\ref{fig:gamma_surf_gamma}a). Both the edge and screw cores dissociate into two partials with a CSF in between, i.e.,
\begin{align}
 \dfrac{1}{2} \langle \bar{1}10]_{\{111\}}= \dfrac{1}{6} \langle\bar{2}11] + \text{CSF}_{\{111\}} + \dfrac{1}{6} \langle \bar{1}2\bar{1}].
\end{align}
The edge core is similar to that in elemental FCC metals; it has a clear planar dissociation with the two partial cores resolved between two close-packed \(\{111\}\) atom-planes and separated by a CSF of width 2.31\(a\). The edge core is thus expected to have relatively low lattice frictions.

The screw \adisl also adopts a planar dissociation, but with a narrow CSF between the two partials. The individual partials are resolved via fractional dislocations with large DD between a pair of Al atoms. This core is different from the nonplanar core calculated using an BOP~\cite{porizek_2002_mrsp,katzarov_2007_pm}. Nonetheless, the BOP has a CSF energy of 412 mJ/m\(^2\), making the dissociation distance less than one lattice spacing and the dissociation impractical. A compact core is also reported in LDA-DFT calculations but the possibility of a dissociated core is not ruled-out~\cite{woodward_2004_pm}. The MTP has accurate CSF energy in agreement with most DFT values (Table~\ref{tab:sf_e}). For the current planar dissociated core in MTP, the partial separation is 1.65a, which is narrower than that in many elemental FCC metals.  The narrow core facilitates the cross-slip and double-cross-slip of the screw segments, leading to frequent jog formations and consistent with a TEM study where the screw dislocation is frequently pinned~\cite{sriram_1997_pma}. Comparisons among all the models suggest that the screw core dissociation is sensitive to the CSF energy, which in turn can be altered by temperature and alloying. The screw core behaviour will thus require further study.

\begin{figure*}[!htbp]
 \centering
 \includegraphics[width=0.70\textwidth]{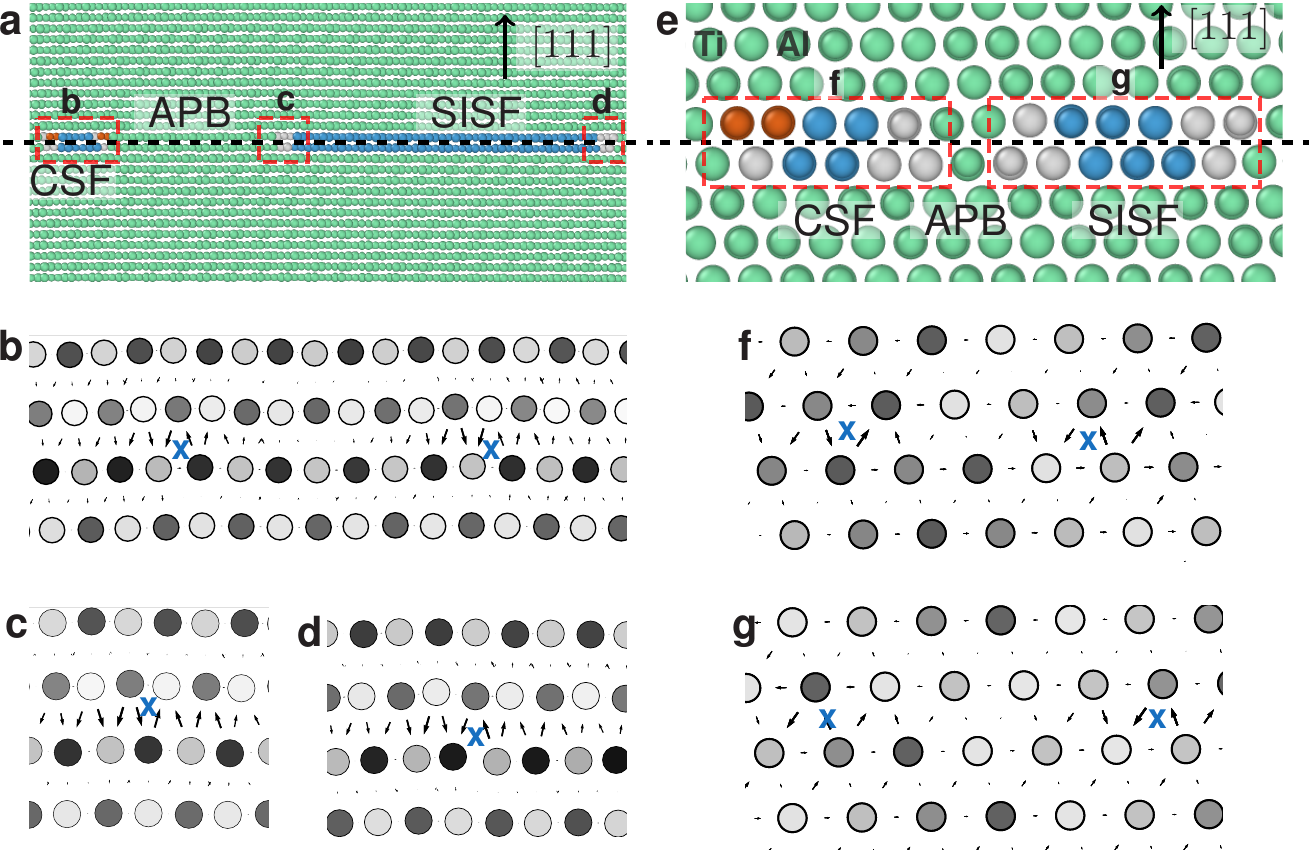}
 \caption{\label{fig:core_sa_gamma} The core structures of the \(\langle 101 ]\) superdislocation on the \{111\} plane in the L1\(_0\) \(\upgamma\)-TiAl structure. (a-d) Edge dislocation core visualized by CNA and DD map. (e-g) Screw dislocation visualized by CNA and DD map. See Fig.~\ref{fig:core_a_gamma} for interpretations of colors. }
\end{figure*}

Figure~\ref{fig:core_sa_gamma} shows the core structure of the \(\langle 101 ]\) super dislocations (Fig.~\ref{fig:gamma_surf_gamma}a). This super dislocation again adopts a planar dissociation into 4 partials as
\begin{equation}
 \begin{aligned}
 \langle 101]_{\{111\}} &= \dfrac{1}{6} \langle 1\bar{1}2] + \text{SISF}_{\{111\}} + \dfrac{1}{6} \langle 211] \\
 &\quad + \text{APB}_{\{111\}} + \dfrac{1}{6} \langle 1\bar{1}2] + \text{CSF}_{\{111\}} + \dfrac{1}{6} \langle 211].
 \end{aligned}
\end{equation}
For the edge core, the partial separations connecting the SISF and APB are particularly wide (16.77\(a\) and 9.47\(a\)), while that connecting the CSF is quite narrow (2.92\(a\)). It is thus possible that the edge core may appear to dissociate into 3 partials in experiments~\cite{hug_1988_pma}. On the other hand, the screw core is much narrower, with the SISF, APB and CSF width at 2.88\(a\), 1.86\(a\) and 2.30\(a\), respectively. The screw core may thus appear to dissociate into 2 partials, as reported in a TEM study~\cite{sriram_1995_msea}.

\begin{figure*}[!htbp]
 \centering
 \includegraphics[width=0.75\textwidth]{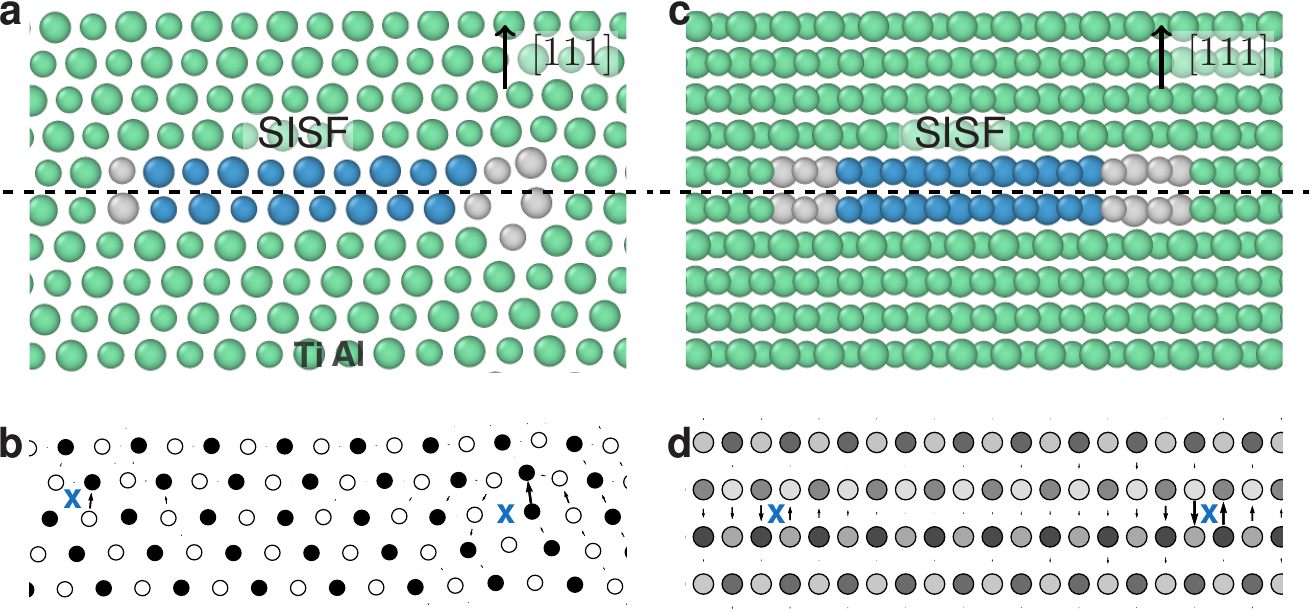}
 \caption{\label{fig:core_sb_gamma} The core structures of the \(\langle 11\bar{2}]/2\) superdislocation on the \{111\} plane in the L1\(_0\) \(\upgamma\)-TiAl structure. (a-b) Edge dislocation core visualized by CNA and DD map. (c-d) Screw dislocation core visualized by CNA and DD map. See Fig.~\ref{fig:core_a_gamma} for interpretations of colors. }
\end{figure*}

Figure~\ref{fig:core_sb_gamma} shows the core structure of the \(\langle 11\bar{2}]/2\) super dislocation on the \(\{111\}\) slip plane (Fig.~\ref{fig:gamma_surf_gamma}a). The super dislocation dissociates into a pair of superpartials as
\begin{equation}
 \dfrac{1}{2}\langle11\bar{2}]_{\{111\}} = \dfrac{1}{6}\langle11\bar{2}] + \text{SISF}_{\{111\}} + \dfrac{1}{3}\langle11\bar{2}]
\end{equation}

The SISF has a width of 6.24\(a\) and 4.93\(a\) for the edge and screw dislocations, respectively. The superpartials have different Burgers vectors and thus different superpartial core structures. These different cores indicate directionality in the glide behaviour of this superdislocation. In addition, the \(\dfrac{1}{3}[11\bar{2}]\) partial core exhibits a tendency of nonplanar dissociation in the edge orientation, which suggests high lattice frictions of the edge segment. The nonplanar dissociation may facilitate further dissociations onto multiple \(\{111\}\) planes as suggested from experimental observation~\cite{kumar_1999_pml}.

\begin{figure*}[!htbp]
 \centering
 \includegraphics[width=0.85\textwidth]{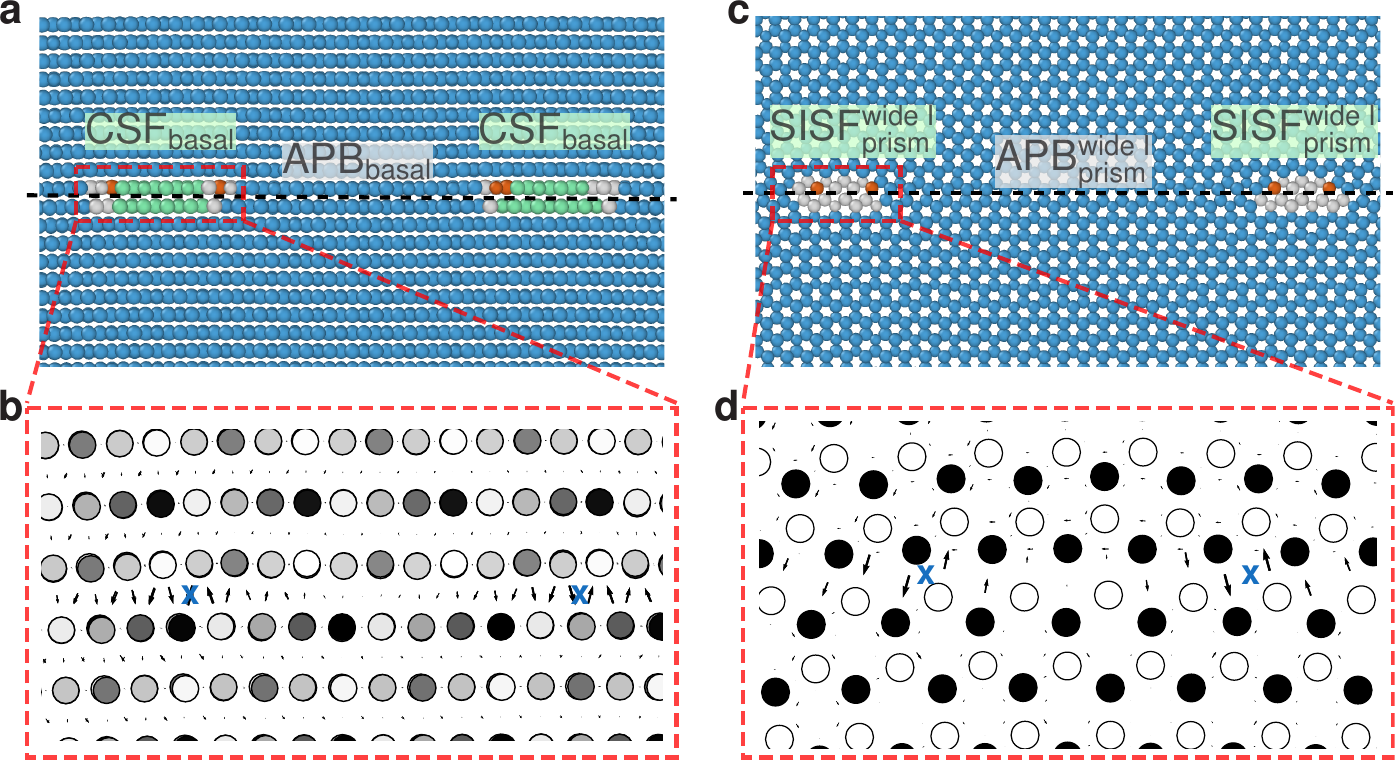}
 \caption{\label{fig:core_edge_a_alpha2} The core structures of the edge \(\langle 11\bar{2}0 ]/3\)-\adisl superdislocation in the D0\(_{19}\) \(\alpha_2\)-\ce{Ti3Al} structure. (a-b) On the \(\{0001\}\) basal plane visualized by CNA and DD map. (c-d) On the \(\{10\bar{1}0\}\) prism wide II plane visualized by CNA and DD map. See Fig.~\ref{fig:core_a_gamma} for interpretations of colors. }
\end{figure*}

In the D0\(_{19}\) \ce{Ti3Al} intermetallic structure, plastic deformation are primarily carried out by the \adisl dislocations on the basal and prism planes and by the \ccadisl dislocations on the pyramidal I and II planes. Figure~\ref{fig:core_edge_a_alpha2} shows the core structures of the edge \(\langle 11\bar{2}0 ]/3\)-\adisl dislocation in the D0\(_{19}\) \(\upalpha_2\)-\ce{Ti3Al} structure. On the basal plane, the edge core dissociates into a pair of superpartials separated by an APB of width 6.04\(a\); the superpartials further dissociate into Shockley partials with a wide CSF of width 2.51\(a\). The total decomposition reaction is
\begin{equation}
 \begin{aligned}
  &\text{Edge }\dfrac{1}{3} \langle 11\bar{2} 0 \rangle_\textbf{basal} = \dfrac{1}{6} \langle 10\bar{1}0 \rangle + \text{CSF}_\textbf{basal} + \dfrac{1}{6} \langle 01\bar{1}0 \rangle\\
  &\quad+ \text{APB}_\textbf{basal} + \dfrac{1}{6} \langle 10\bar{1}0 \rangle + \text{CSF}_\textbf{basal} + \dfrac{1}{6} \langle 01\bar{1}0 \rangle
 \end{aligned}
\end{equation}
The \adisl dislocation may thus appear as 4 partials in TEM observations. Furthermore, all the partial cores are resolved between two close-packed \(\{0001\}\) atom-planes and have planar dissociations, as seen in the DD map (Fig.~\ref{fig:core_edge_a_alpha2}b). The edge \adisl is thus expected to have relatively low lattice frictions.

On the prism wide II plane (Fig.~\ref{fig:xtal_plane}c-d), the edge core dissociates into a pair of superpartials separated by a wide APB of 10.12\(a\) on the prism plane; the individual superpartials further dissociate into a pair of partials with a narrow SISF of 1.88\(a\) in between, which is similar to that in HCP Ti. The narrow superpartial core is consistent with the high SF energy on this slip plane (Fig.~\ref{fig:gamma_surf_prism_alpha2}k). The total decomposition reaction is
\begin{equation}\label{eq:edge_a_disso_alpha2}
\begin{aligned}
 &\text{Edge }\dfrac{1}{3} \langle 11\bar{2} 0 \rangle_\textbf{prism}^{\text{wide II}} = \dfrac{1}{12} \langle 11\bar{2}X \rangle + \text{SISF}^{\text{wide II}}_\textbf{prism} + \dfrac{1}{12} \langle 11\bar{2} \bar{X} \rangle\\
 &\quad+ \text{APB}^{\text{wide II}}_\textbf{prism} + \dfrac{1}{12} \langle 11\bar{2}X \rangle + \text{SISF}^{\text{wide II}}_\textbf{prism} + \dfrac{1}{12} \langle 11\bar{2} \bar{X} \rangle
\end{aligned}
\end{equation}
where \(X\) accounts for some possible components in the crystallographic \cdisl direction as suggested by the \(\gamma\)-surfaces of this plane (Fig.~\ref{fig:gamma_surf_prism_alpha2}d-f). Compared to the dissociation on the close-packed basal plane (Fig.~\ref{fig:core_edge_a_alpha2}a-b), the Burgers vector of the superpartials is resolved through DD on 4 \(\{10\bar{1}0\}\) atom-planes on this prism plane. Glide of the edge \adisl dislocations may thus require higher stresses relative to that on the planar dissociated basal plane. Furthermore, given the narrow superpartial cores, the \adisl core may thus appear to decompose into two superpartials in TEM study~\cite{wiezorek_1995_pml}.

\begin{figure*}[!htbp]
 \centering
 \includegraphics[width=0.80\textwidth]{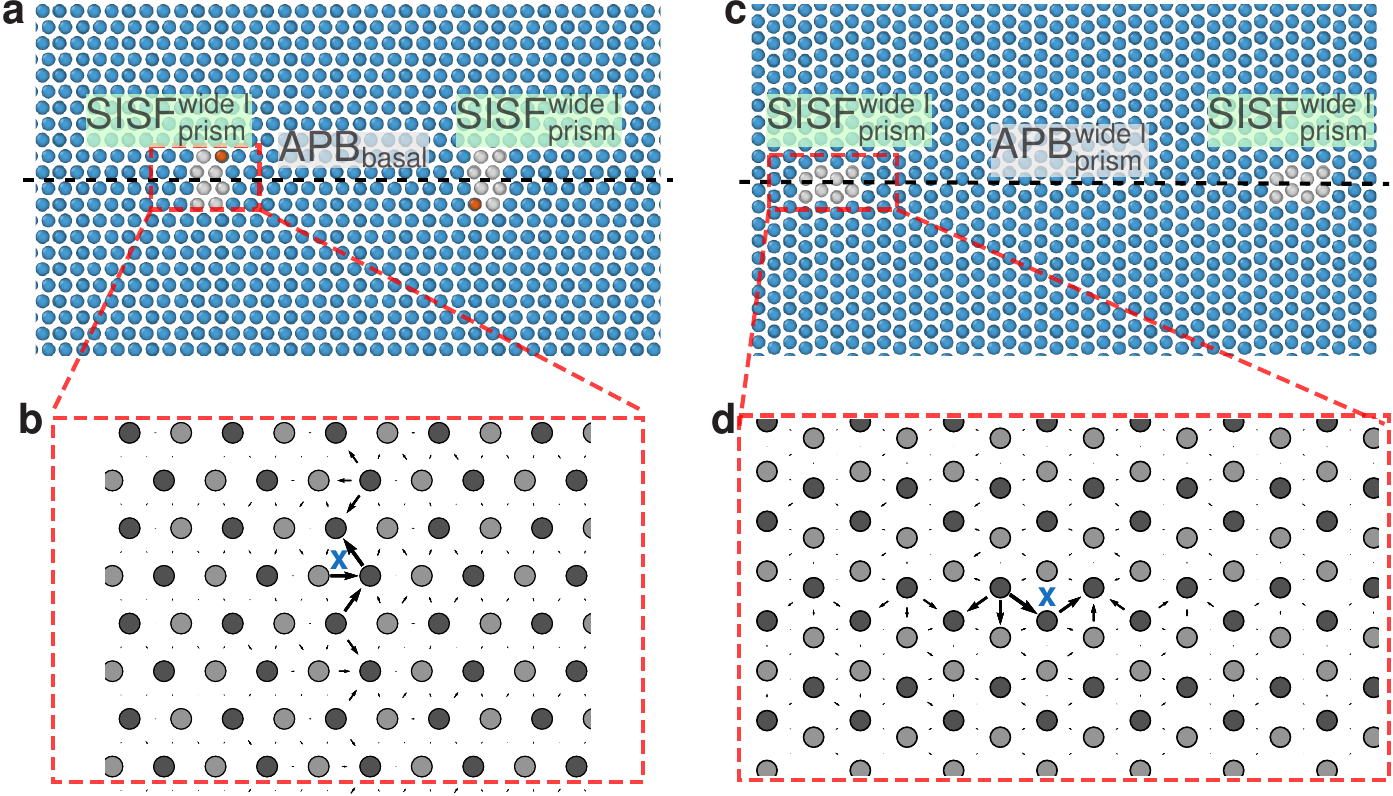}
 \caption{\label{fig:core_screw_a_wide_ii_alpha2} The core structures of the screw \(\langle 11\bar{2}0 ]/3\)-\adisl superdislocation in the D0\(_{19}\) \(\alpha_2\)-\ce{Ti3Al} structure. (a-b) On the \(\{0001\}\) basal plane visualized by CNA and DD map. (c-d) On the \(\{10\bar{1}0\}\) prism wide II plane visualized by CNA and DD map. See Fig.~\ref{fig:core_a_gamma} for interpretations of colors. }
\end{figure*}

Figure~\ref{fig:core_screw_a_wide_ii_alpha2} shows the core structures of the screw \(\langle 11\bar{2}0 ]/3\)-\adisl superdislocation in the D0\(_{19}\) \(\upalpha_2\)-\ce{Ti3Al} structure. On the basal plane, the screw core dissociates into a pair of superpartials separated by an APB of width 6.47\(a\) on the basal plane; the superpartials further dissociate into Shockley partials, which spontaneously cross-slip to and dissociate into partials on the prism wide II plane. The screw \adisl core thus has a non-coplanar structure with an APB on the basal plane and superpartial cores along the prism plane. The total decomposition reaction is
\begin{equation}
 \begin{aligned}
  &\text{Screw } \dfrac{1}{3} \langle 11\bar{2} 0 \rangle_\textbf{basal} = \dfrac{1}{12} \langle 11\bar{2}X \rangle + \text{SISF}_\textbf{prism}^\text{wide II} + \dfrac{1}{12} \langle 11\bar{2}\bar{X} \rangle \\
  &\quad+ \text{APB}_\textbf{basal} + \dfrac{1}{12} \langle 11\bar{2}X \rangle + \text{SISF}_\textbf{prism}^\text{wide II} + \dfrac{1}{12} \langle 11\bar{2}\bar{X} \rangle
 \end{aligned}
\end{equation}

Dissociation of the superpartials of the basal \adisl onto the prism plane is not surprising, given their pure screw character. However, the SISF energy on the prism plane appears to be 25\% higher than the CSF energy on the basal plane, which suggests further multi-layer atomic relaxation in the SISF on the prism wide II plane. This non-coplanar dissociated screw core is also reported previously using a Finnis-Sinclair IAP~\cite{cserti_1992_msea}; this dissociation is expected to have very high lattice friction, consistent with experimental observations of straight segments of the \adisl dislocation in the screw orientation~\cite{legros_1996_pma}. The non-glide of the screw segments makes basal \adisl slip difficult in the \(\upalpha_2\)-\ce{Ti3Al} structure as observed in experiments~\cite{inui_1993_pma}, even though the basal plane has relatively low GSFE and thus favourable slip conditions (Fig.~\ref{fig:gamma_surf_basal_alpha2}).

On the prism wide II plane, the screw \adisl core dissociates into a pair of superpartials separated by an APB\(_\text{prism}^\text{wide II}\) of width 10.89\(a\) (Fig.~\ref{fig:core_screw_a_wide_ii_alpha2}c-d); each superpartial further dissociates into a pair of narrowly-separated Shockley partials. The superpartials have core structures similar to that of the screw \adisl in HCP Ti~\cite{wen_2021_npjcm}. The total decomposition reaction is
\begin{equation}\label{eq:screw_a_disso_wide_ii_alpha2}
 \begin{aligned}
 &\text{Screw } \dfrac{1}{3} \langle 11\bar{2} 0 \rangle_\textbf{prism}^\text{wide II} = \dfrac{1}{12} \langle 11\bar{2}X \rangle + \text{SISF}^\text{wide II}_{\textbf{prism}} + \dfrac{1}{12} \langle 11\bar{2} \bar{X} \rangle\\
  &\quad+ \text{APB}_{\textbf{prism}}^\text{wide II} + \dfrac{1}{12} \langle 11\bar{2}X \rangle + \text{SISF}_\textbf{prism}^\text{wide II} + \dfrac{1}{12} \langle 11\bar{2} \bar{X} \rangle.
 \end{aligned}
\end{equation}
The co-planar dissociation suggests the super screw \adisl dislocations have a relatively low lattice friction on the prism plane than on the basal plane. The prism wide II plane also has a relative low energy \(\gamma\)-lines in the \adisl direction (Fig.~\ref{fig:gamma_surf_prism_alpha2}k). Therefore, the prism wide II plane is expected to be the primary slip plane for \adisl slip in \(\alpha_2\)-\ce{Ti3Al}.

\begin{figure*}[!htbp]
 \centering
 \includegraphics[width=0.95\textwidth]{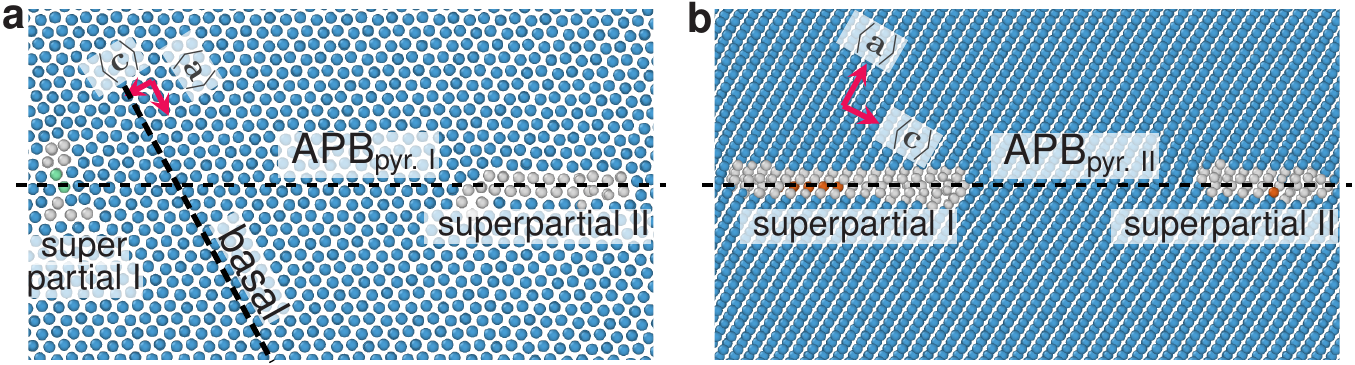}
 \caption{\label{fig:core_mix_edge_cca_alpha2} The core structures of the mixed and edge \ccadisl superdislocation in the D0\(_{19}\) \(\alpha_2\)-\ce{Ti3Al} structure. (a) Mixed \ccadisl on pyramidal I plane visualized by CNA. (b) Edge \ccadisl on pyramidal II plane visualized by CNA. In (a), the left superpartial of the mixed \ccadisl has a climb-dissociated core configuration along the basal plane. See Fig.~\ref{fig:core_a_gamma} for interpretations of colors. }
\end{figure*}

Figure~\ref{fig:core_mix_edge_cca_alpha2} shows the core structures of the mixed and edge \ccadisl dislocations dissociated on pyramidal I and II planes, respectively. On the pyramidal I plane, the dislocation first dissociates into two superpartials separated by an APB; the two superpartials further dissociate into a pair of partial dislocations and separated by a CSF in between, consistent with the pyramidal I plane \(\gamma\)-surface (Fig.~\ref{fig:gamma_surf_pyi_alpha2}). The total decomposition is can be approximately expressed as
\begin{equation}
\begin{aligned}
 &\text{Mixed} \dfrac{1}{3}\langle 11\bar{2} 6 \rangle = Y\langle 11\bar{2} 6 \rangle + \text{CSF}^\text{wide}_\textbf{pyr. I} + (\dfrac{1}{6}-Y)\langle 11\bar{2} 6 \rangle \\
 &\quad + \text{APB}^\text{wide}_\textbf{pyr. I} + Z\langle 11\bar{2} 6 \rangle + \text{CSF2}^\text{wide}_\textbf{pyr. I} + (\dfrac{1}{6}-Z)\langle 11\bar{2} 6 \rangle
\end{aligned}
\end{equation}
where \(Y\) and \(Z\) can not be determined based on crystal symmetry and thus depend on the profiles of the pyramidal I plane \(\gamma\)-surface. Furthermore, the individual superpartial Burgers vectors may have some small components perpendicular to the \ccadisl direction, since the meta-stable CSF positions are not along the \ccadisl path. The exact Burgers vectors are determined by minimizing the sum of the CSF energy, near-core energy and the elastic energy~\cite{yin_2017_am}. The individual superpartials have core structures similar to the \cadisl dislocations in elemental HCP metals such as Mg and Ti (c.f., Fig~\ref{fig:core_mix_edge_cca_alpha2}b and Fig.~4 of Ref.~\cite{wu_2016_sm}). More importantly, the superpartials are not stable and undergo a pyramidal to basal (PB) climb-dissociation, (e.g., the left super partial in Fig.~\ref{fig:core_mix_edge_cca_alpha2}a), again similar to that in Mg or Ti. The climb dissociated superpartials are sessile, as the the stacking fault is in the non-glide basal plane.

On the pyramidal II plane, the edge \ccadisl dislocation dissociates into a pair of superpartials with an APB in between. The individual superpartials further dissociate into two partial dislocations, similar to that on the pyramidal I plane. All partial dislocations are of pure edge character, since all CSF and APB lie along the \ccadisl path. The total decomposition can be written as
\begin{equation}
 \begin{aligned}
  & \dfrac{1}{3}\langle 11\bar{2} 6 \rangle_\textbf{pyr. II} = Y\langle 11\bar{2} 6 \rangle + \text{CSF}_\textbf{pyr. II} + (\dfrac{1}{6}-Y)\langle 11\bar{2} 6 \rangle \\
  &\quad + \text{APB}_\textbf{pyr. II} + Z\langle 11\bar{2} 6 \rangle + \text{CSF2}_\textbf{pyr. II} + (\dfrac{1}{6}-Z)\langle 11\bar{2} 6 \rangle
 \end{aligned}
\end{equation}
The core structures of the two superpartials are again similar to that in Mg and Ti, and are thus susceptible to the PB transition.

\begin{figure*}[!htbp]
 \centering
 \includegraphics[width=0.80\textwidth]{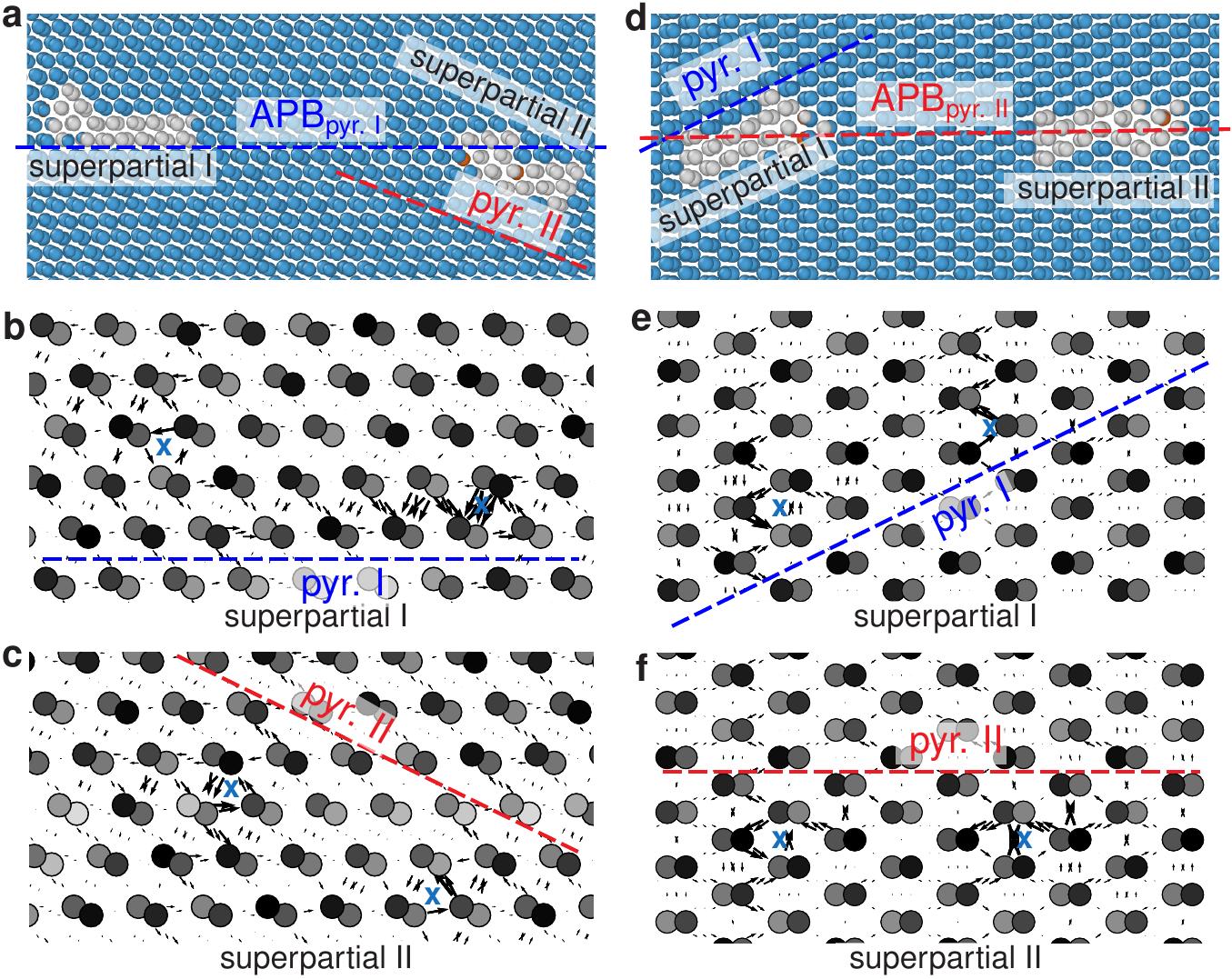}
 \caption{\label{fig:core_screw_cca_alpha2} The core structures of the screw \ccadisl super dislocation in the D0\(_{19}\) \(\alpha_2\)-\ce{Ti3Al} structure. (a) Screw \ccadisl on the Pyramidal I plane visualized by CNA. (b-c) Core structures of the two superpartials in (a) visualized by DD map. (d) Screw \ccadisl on the Pyramidal II plane visualized by CNA. (e-f) Core structures of the two superpartials in (d) visualized by DD map. See Fig.~\ref{fig:core_a_gamma} for interpretations of colors. }
\end{figure*}

Figure~\ref{fig:core_screw_cca_alpha2} shows the core structures of the screw \ccadisl dislocations dissociated on pyramidal I and II planes. The dissociations are largely similar to that of the mixed and edge \ccadislns; the dislocation has a pair of superpartials with a wide APB of 6.22\(a\)) and 4.89\(a\) on the pyramidal I or II slip plane and the superpartials further dissociate into two partials with a core structure similar to that in elemental Mg and Ti (c.f. Fig.~\ref{fig:core_screw_cca_alpha2} and Fig.~3 of Ref.~\cite{itakura_2016_prl}). All partials are of nearly screw character. The total decomposition can be written as
\begin{equation}
 \begin{aligned}
  & \dfrac{1}{3}\langle 11\bar{2} 6 \rangle_\textbf{pyr. I/II} = Y\langle 11\bar{2} 6 \rangle + \text{CSF}_\textbf{pyr. I/II} + (\dfrac{1}{6}-Y)\langle 11\bar{2} 6 \rangle \\
  &\ + \text{APB}_\textbf{pyr. I/II} + Z\langle 11\bar{2} 6 \rangle + \text{CSF2}_\textbf{pyr. II/I} + (\dfrac{1}{6}-Z)\langle 11\bar{2} 6 \rangle.
 \end{aligned}
\end{equation}

Nonetheless, the screw \ccadisl dissociation is unique as the two superpartials always lie on different pyramidal planes, regardless of the APB plane. The nonplanar dissociation is governed by the competition of the 4 CSFs on the pyramidal I and II planes (c.f. Fig.~\ref{fig:gamma_surf_pyi_alpha2}f and Fig.~\ref{fig:gamma_surf_pyii_alpha2}f). This competition is delicate; the MLP3 overestimates the CSFs by 5-10\% and the classical IAPs are qualitatively different \(\gamma\)-lines, while only the MTP has the respective CSFs in agreement with DFT. Since the MTP has accurate CSF energies, the surprising dissociation may thus be intrinsic in the D0\(_{19}\)-\ce{Ti3Al} structure.

\section{Discussion}
\label{sec:disc}

The above comprehensive benchmarks show that the current machine-learning (ML) MTP reproduces both the bulk and defect properties largely in agreement with experiments and/or DFT. The MTP has exceptional accuracies in many properties relevant to plastic deformation in both the L1\(_0\) and D0\(_{19}\) phases. In particular, the MTP quantitatively reproduces nearly all the \(\gamma\)-lines from DFT, even though none of the stacking fault structures is explicitly contained in the training datasets. The MLP3 also exhibits \(\gamma\)-line profiles qualitatively similar to that of the MTP and DFT, despite using a different ML framework. This suggests some transferability and robustness of ML approaches in developing IAPs in the TiAl system. Both ML-IAPs are generally more accurate in describing the alloy properties than the IAPs based on semi-empirical classical approaches. The classical IAPs also have serious shortcomings in describing the GSFEs in the D0\(_{19}\) phase. Apart from the differences in the IAP formalisms, the fitting procedures used in the individual approaches may have contributed to the differences in accuracies in respective structures. The classical IAPs are developed based on potentials for elemental HCP Ti and FCC Al, while the ML-IAPs are trained using datasets consisting of a large class of unary and binary structures from the beginning. In the classical IAPs, further improvements may still be possible if they are developed with a balance of all relevant structures.

In the ML-IAPs, quantitative accuracy is not guaranteed either. The MLP3 is trained based on bulk structures and their distorted variations. It exhibits considerable discrepancies in surface energies in all the structures examined here and in the metastable SF energies such as the various APBs in the D0\(_{19}\) phase. The MTP has much-improved accuracy in both surface and SF energies, which may be due to the inclusion of surface structures in the MTP training datasets. The inclusion of special defect structures is also necessary even in training ML-IAPs for elemental metals such as Ti~\cite{wen_2021_npjcm} and V~\cite{wang_2022_prm}. The current work suggests that ML frameworks have the flexibility for continuous improvements when more relevant configurations are incorporated in the training datasets. For modelling crystal defects such as dislocations which are generally more sensitive to model parameters, ML-IAPs should be quantitatively benchmarked with DFT to validate their appropriateness.

In the preliminary study on dislocations, the current MTP reveals unique core structures in both intermetallic phases. In the L1\(_0\) structure, the ordinary screw \adisl dislocation has a planar dissociation with a CSF of only 1.65\(a\) in between the two partials, in contrast to the non-planar core in LDA-DFT~\cite{woodward_2004_pm} and BOP~\cite{porizek_2002_mrsp,znam_2003_pm}. Since the CSF width is narrow and similar to the SF width in the screw \adisl in FCC Al~\cite{woodward_2008_prl}, the planar dissociation can be sensitive to the CSF energy and elastic constants, which in turn can be influenced by temperature and alloy compositions. The BOP predicts a CSF of 412 mJ/m\(^2\) higher than most DFT values (Table~\ref{tab:sf_e}). The exact core structure may thus require further study, in both DFT with GGA functionals and MD at finite temperatures. Nonetheless, the narrowly-dissociated core suggests its easy cross-slip, either intrinsic or extrinsic, which is consistent with broad experimental observations~\cite{sriram_1997_pma,zghal_1998_am,couret_1999_pma,katzarov_2011_am}. The screw \(\langle 101 ]\) superdislocation also has a narrow CSF bounded by two partial cores (Fig.~\ref{fig:core_sa_gamma}e). This superdislocation may thus cross-slip onto other \(\{111\}\) slip planes under thermal activation, as observed in MD simulations using the classical IAPs and in an early study using a BOP~\cite{porizek_2002_mrsp,katzarov_2009_pm}.

In the D0\(_{19}\) structure, all superdislocations have extraordinarily wide dissociations, making first-principles DFT calculations of core structures impractical. Since the MTP has accurate \(\gamma\)-lines on all relevant slip planes, the core structures predicted here should be reliable. In particular, the basal plane screw \adisl superdislocation dissociation is unconventional with its partial cores spreading on the prism plane and an APB on the basal plane (Fig.~\ref{fig:core_screw_a_wide_ii_alpha2}a). This core is sessile and in a locked configuration, which explains the much higher CRSS for basal plane slip than that for prism plane slip, even though the basal plane has a generally lower \(\gamma\)-line in the slip direction. The screw \ccadisl superdislocation cores are also unusual (Fig.~\ref{fig:core_screw_cca_alpha2}). With their two superpartials on different pyramidal planes, the glides of the \ccadisl superdislocations are beyond established models and can only be determined in molecular static or dynamic simulations. Furthermore, these peculiar dissociations appear to be governed by the SF energies on competing slip planes, and thus may be changed by appropriate solid solution alloying. The current study provides the physical basis for further DFT calculations in controlling SF energy and ultimately achieving planar core dissociations in the D0\(_{19}\) structure. The stability of the mixed and edge \ccadisl superdislocations may be the key barrier in improving the ductility of the D0\(_{19}\) phase. With the MTP, the kinetic barrier for the PB transformation can be determined, similar to that in elemental Mg~\cite{wu_2015_nature}. The effects of alloy compositions may also influence the PB transition and help stabilize these superdislocations at higher solute concentrations, a scenario possibly achievable in compositionally complex alloys.

\section{Conclusions}
\label{sec:conc}

In summary, we have developed an interatomic potential for the Ti-Al binary system using the machine learning moment tensor potential framework. This potential is fitted with a wide range of datasets based on bulk and defect structures in the Ti-Al system. The resulting potential accurately reproduces bulk lattice and defect properties in both the L1\(_0\) \(\gamma\)-TiAl and D0\(_{19}\) \(\alpha_2\)-\ce{Ti3Al} intermetallic structures. The overall accuracy is much improved compared to previous machine-learning and classical interatomic potentials. Preliminary molecular static simulations reveal all relevant dislocation core structures largely consistent with the generalized stacking fault energies from DFT and extant observations in experiments.  The developed potential combined with DFT-based \(\gamma\)-lines also shed lights on the complex \adisl and \ccadisl slip systems in the D0\(_{19}\) \(\alpha_2\)-\ce{Ti3Al} structure.  While this potential may still be fine-tuned for the screw \adisl ordinary dislocation in the L1\(_0\) phase,  the present version enables quantitative modelling of dislocation and fracture properties in a self-consistent manner in both phases and perhaps at the two-phase interfaces. For examples, the critical resolved shear stresses and mobilities of individual dislocations can be determined using molecular static and dynamics simulations. In addition, the new potential makes it possible to study non-Schmid effects such as slip directionality, thermally activated dislocation cross-slip and non-planar climb dissociations in both phases. The current work also provides a comprehensive datasets for future benchmark, development and fine-tuning of interatomic potentials in binary Ti-Al and higher order Ti-Al-based intermetallic systems.

\section{Acknowledgements}

SSQ, XQP, ZA, AMZT and MHJ acknowledge support from A*STAR, Singapore via the Structural Metals and Alloys Programme (Grant No.: A18B1b0061).  TW acknowledges support by the University of Hong Kong (HKU) via a seed fund (2201100392).  ZW acknowledges financial support and computational resources from City University of Hong Kong (Grants No.: 7005454 \& 9610436).  The computational work for this article was performed on Expanse at San Diego Supercomputer Center (SDSC) from the Advanced Cyberinfrastructure Coordination Ecosystem: Services \& Support (ACCESS) program supported by National Science Foundation (Grants No.: 2138259, 2138286, 2138307, 2137603, and 2138296) and the National Supercomputing Centre (NSCC), Singapore (https://www.nscc.sg).

\end{document}